\RequirePackage[l2tabu,orthodox]{nag}
\documentclass[copyright,creativecomons,noderivs,10pt]{eptcs}

\providecommand{\event}{QPL 2017}

\pdfoutput=1

\usepackage{rwd-drafting}

\usepackage[T1]{fontenc}
\usepackage{lmodern}
\usepackage{fixltx2e}
\usepackage{microtype} 
\usepackage{xspace}

\newcommand{\figureline}{\rule{\textwidth}{0.5pt}}

\makeatletter
\newcommand\etc{etc\@ifnextchar.{}{.\@}\xspace}
\newcommand\ie{i.e.\@\xspace}  
\newcommand\eg{e.g.\@\xspace}
\makeatother

\usepackage{graphicx}

\newcommand{\inlinegraphic}[2]{
  \dimendef\grafheight=255\dimendef\grafvshift=254
  \grafheight=#1
  \grafvshift=-0.5\grafheight
  \advance\grafvshift by 0.5ex
  \raisebox{\grafvshift}{\includegraphics[height=\grafheight]{images/#2}\xspace}
}

\newcommand{\ninlinegraphic}[2][1.0]{
  \dimendef\grafheight=255\dimendef\grafvshift=254
  \setbox0 = \hbox{\scalebox{#1}{\includegraphics{images/#2}}}
  \grafheight=\the\ht0
  \grafvshift=-0.5\grafheight
  \advance\grafvshift by 0.5ex
  \raisebox{\grafvshift}{\includegraphics[height=\grafheight]{images/#2}\xspace}
}

\newcommand{\inline}[1]{
  \raisebox{0.5ex}{\;#1\;}
}

\usepackage{latexsym}
\usepackage{amssymb}
\usepackage{amsmath} 
\usepackage{stmaryrd}
\usepackage{mathtools}

\usepackage{amsthm}
\newtheorem{theorem}{Theorem}[section]
\newtheorem{proposition}[theorem]{Proposition}

\newtheorem{corollary}[theorem]{Corollary}
\theoremstyle{definition}
\theoremstyle{definition}
\theoremstyle{definition}\newtheorem{definition}[theorem]{Definition}
\theoremstyle{definition}
\theoremstyle{definition}
\theoremstyle{definition}

\newcommand{\denote}[1]{
\llbracket #1 \rrbracket} 
\newcommand{\ldenote}[1]{\left\llbracket #1 \right\rrbracket}

\newcommand{\ket}[1]{
    \ensuremath{\left|  #1 \right\rangle}\xspace}

\usepackage{diagrams} 
\newarrow{Equals}{}{=}{}{=}{}

\ifx\numericids\undefined
\newcommand{\id}[1]{\ensuremath{\mathrm{id}_{#1}}}
\else
\newcommand{\id}[1]{\ensuremath{1_{#1}}}
\fi

\newcommand{\fdhilb}{
\ensuremath{\mathbf{fdHilb}}\xspace}

\newcommand{\zxcalculus}{\textsc{zx}-calculus\xspace}

\usepackage{dsfont}

\newcommand{\Enc}{\ensuremath{\mathrm{Enc}}\xspace}
\newcommand{\Dec}{\ensuremath{\mathrm{Dec}}\xspace}

\usepackage{hyperref}

\usepackage{tikz}
\usepackage{pgfplots}
\usetikzlibrary{trees}
\usetikzlibrary{topaths}
\usetikzlibrary{decorations.pathmorphing}
\usetikzlibrary{decorations.markings}
\usetikzlibrary{matrix,backgrounds,folding}
\usetikzlibrary{chains,scopes,positioning,fit}
\usetikzlibrary{arrows,shadows}
\usetikzlibrary{calc} 
\usetikzlibrary{chains}
\usetikzlibrary{shapes,shapes.geometric,shapes.misc}

\usepackage[undirected]{tikzquanto}
\usepackage{tikzcircuits}

\pgfdeclarelayer{edgelayer}
\pgfdeclarelayer{nodelayer}
\pgfsetlayers{background,edgelayer,nodelayer,main}

\tikzstyle{every picture}=[baseline=(current bounding box).east,node distance=5mm]
\tikzstyle{none}=[inner sep=0mm]
\tikzstyle{every loop}=[]

\tikzstyle{(null)}=[]
\tikzstyle{plain}=[]

\tikzset{
operator/.style = {draw,fill=white,minimum size=1.5em},%
operator2/.style ={draw,fill=white,minimum height=7cm,minimum width=1.5 cm},%
phase/.style = {draw,fill,shape=circle,minimum size=5pt,inner sep=0pt},%
endline/.style = {draw,fill,shape=circle,minimum size=1pt,inner sep=0pt},%
surround/.style = {fill=blue!10,thick,draw=black,rounded corners=2mm},%
cross/.style={},
circlewc/.style={draw,circle,cross,minimum width=0.3 cm},%
} 

\newcommand{\circbox}[1]{\inline{%
\begin{tikzpicture}[circuit]
  \node  (0) at (1,0) {};
  \node [box] (1) at (0, 0) {$#1$};
  \node  (2) at (-1,0) {};
  \draw  (0) to (1);
  \draw  (1) to (2);
\end{tikzpicture}
}}

\newcommand{\circX}[1]{\circbox{X_#1}}
\newcommand{\circZ}[1]{\circbox{Z_#1}}
\newcommand{\circH}{\circbox{H}}

\newcommand{\circCX}{\inline{%
\begin{tikzpicture}[circuit]
  \node  (0) at (1, 0) {};
  \node  (1) at (1, 1) {};
  \circcnot{3}{0,1}{2}{0,0}
  \node  (4) at (-1, 0) {};
  \node  (5) at (-1,1) {};
  \draw  (1) to (3);
  \draw  (0) to (2);
  \draw  (3) to (5);
  \draw  (2) to (4);
\end{tikzpicture}
}}

\newcommand{\greenunit}{\inline{%
  \begin{tikzpicture}[quanto]
    \node [boundary vertex] (a) at (2.0,1.0) {};
    \node [green vertex] (b) at (1.0,1.0) {};
    \draw [] (b) to (a);
  \end{tikzpicture}
}}

\newcommand{\greenpoint}[1]{\inline{%
  \begin{tikzpicture}[quanto]
    \node [boundary vertex] (a) at (2.0,-1.0) {};
    \node [green vertex] (b) at (1.0,-1.0) {};
    \draw [] (b) to (a);
    \node [green angle left] at (b) {$#1$};
  \end{tikzpicture}
}}

\newcommand{\redpoint}[1]{\inline{%
  \begin{tikzpicture}[quanto]
    \node [boundary vertex] (a) at (2.0,-1.0) {};
    \node [red vertex] (b) at (1.0,-1.0) {};
    \draw [] (b) to (a);
    \node [red angle left] at (b) {$#1$};
  \end{tikzpicture}
}}

\newcommand{\redunit}{\inline{%
  \begin{tikzpicture}[quanto]
    \node [boundary vertex] (a) at (2.0,1.0) {};
    \node [red vertex] (b) at (1.0,1.0) {};
    \draw [] (b) to (a);
  \end{tikzpicture}
}}

\usepackage{xr-hyper}
\newtheorem{lem}{Lemma}

\title{Verifying the Smallest Interesting Colour Code with Quantomatic}
\author{Liam Garvie and Ross Duncan
\institute{University of Strathclyde \\ 26 Richmond Street, Glasgow G1 1XH, Scotland}
}
\date{\today}

\def\titlerunning{The Smallest Interesting Colour Code in Quantomatic}
\def\authorrunning{L. Garvie \& R. Duncan}

\begin{document}
\maketitle
\begin{abstract}
  In this paper we present a Quantomatic case study, verifying the
  basic properties of the \emph{Smallest Interesting Colour Code}
  error detection code.
\end{abstract}

\section{Introduction}
\label{sec:introduction}


Error correction will form a crucial layer in the software
stack of any realistic quantum computer for the foreseeable
future.  Since the implementation of any error correction scheme will
depend on the details of the actual hardware, it is generally expected
that error correction will be added to a quantum program late in the
compilation process (see, for example \cite{Hner2016A-Software-Meth}).
In other words, the fault-tolerant executable program will be
\emph{automatically generated} from a higher-level description which
is unaware of the error-correction scheme to be employed.
%
%
At a minimum, any such translation process from ``logical'' quantum
circuits to their fault-tolerant versions should be proven
\emph{sound}---\ie that translation does not change the meaning of the
program.  However we might demand more.  For example, with knowledge of the
hardware operations and the fault-tolerant program, it
may be possible to optimise beneath the error correction scheme.  Such
optimisations again require correctness proofs, not just of soundness,
but to guarantee that the optimisation preserves fault-tolerance.

To do any of this, a language combining circuits, error-correcting
schemes, and translations between the two is required, and this
language should support robust and powerful automated reasoning,
capable of proving that relevant properties of error-correcting
schemes hold.  In this paper we present a case study along these
lines.  Precisely, we use the \zxcalculus \cite{Coecke:2009aa} as a
language, in concert with the interactive theorem prover Quantomatic
\cite{quantomatichome,Kissinger2015Quantomatic:-A-}, to study the
\emph{Smallest Interesting Colour Code} \cite{Campbell:colourcode}.
We provide formal proofs of the basic properties of the code itself
and its fault-tolerant operations.

A similar study was conducted in 2013 for the
7-qubit Steane code \cite{Duncan:2013lr} (see also the recent
  \cite{Chancellor2016Coherent-Parity}), however the
Quantomatic system has undergone significant development in the
intervening four years, and is vastly more powerful. Many of our
proofs can be produced automatically by Quantomatic; others require
the formalisation of human insight into reusable \emph{tactics}; still
others resist full automation with the current technology.  In the final
section we discuss the obstacles encountered, and desiderata for
future development of the \zxcalculus / Quantomatic system.


\paragraph{Acknowledgements}
\label{sec:acknowledgements}

The authors wish to thank Aleks Kissinger:
without his timely intervention this project would have never have
been completed.

\paragraph{The Quantomatic project files}
\label{sec:quant-proj-files}
All the proofs which appear in this paper and its appendix are
publicly available as a downloadable Quantomatic project at 
\url{https://gitlab.cis.strath.ac.uk/kwb13215/Colour-Code-QPL}.

\paragraph{Supplementary Material}
\label{sec:suppl-mater}

Links to the supplementary material, containing the full proofs of the
main results are found in Appendix \ref{sec:proofs-omitted-from}.

\section{The Code}
\label{sec:code}

The ``smallest interesting colour code'' takes its name from a blog
post by Earl Campbell \cite{Campbell:colourcode} which provided the
inspiration for this project, and also appears in the papers
\cite{Campbell2016Unifying-gate-s,PhysRevA.95.022316,PhysRevA.95.032338}.
It is an [[8, 3, 2]] code, meaning it encodes 3 logical qubits into 8
physical qubits, and has a distance of 2, meaning it can detect any
single qubit error but not correct it.  The code can be presented
geometrically as a cube where each vertex corresponds to a qubit.
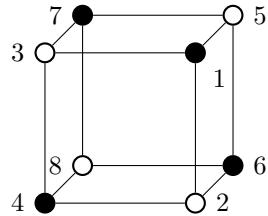
\ctikzfig{cube} 
In this picture, the $Z$-stabilizers of the code
correspond to the face and cell operators of the cube, and the single
$X$-stabilizer corresponds to the cell; see \cite{Campbell:colourcode}
for nice pictures.

The logical Pauli operators are transversal, with the three logical
$X$ operators obtained by applying $X$ to the three faces of the cube,
while the logical $Z$ is obtained as an edge operator.

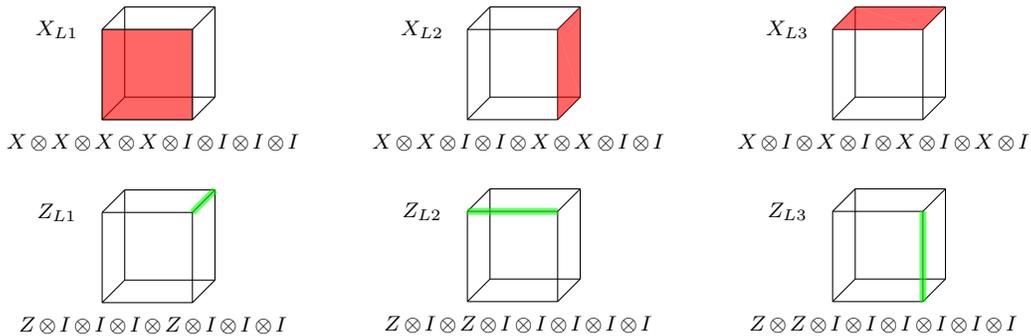
\begin{center}
  \begin{tabular}{ccccc}
\\
\input{figures/logical-X.tikz}  & \qquad &
\input{figures/logical-X2.tikz} & \qquad &
\input{figures/logical-X3.tikz}
\\
\\    
\input{figures/logical-Z.tikz}  & \qquad &
\input{figures/logical-Z2.tikz} & \qquad &
\input{figures/logical-Z3.tikz}
\\
\\
  \end{tabular}
\end{center}

\noindent
From the description of the $X$ operators, it's straightforward to
find the translation of the computational basis states to
codewords, which is shown in Table \ref{tab:basis}.  

\begin{table}
  \centering
  \begin{tabular}{|c|c|}
    logical state & codeword state\\\hline
    $\ket{000}$ & $\frac{1}{\sqrt{2}}(\ket{00000000} + \ket{11111111}) $\\
    $\ket{001}$ & $\frac{1}{\sqrt{2}}(\ket{10101010} + \ket{01010101}) $\\
    $\ket{010}$ & $\frac{1}{\sqrt{2}}(\ket{11001100} + \ket{00110011}) $\\
    $\ket{011}$ & $\frac{1}{\sqrt{2}}(\ket{01100110} + \ket{10011001}) $\\
  \end{tabular}
  \begin{tabular}{|c|c|}
    logical state & codeword state\\\hline
    $\ket{100}$ & $\frac{1}{\sqrt{2}}(\ket{11110000} + \ket{00001111}) $\\
    $\ket{101}$ & $\frac{1}{\sqrt{2}}(\ket{01011010} + \ket{10100101}) $\\
    $\ket{110}$ & $\frac{1}{\sqrt{2}}(\ket{00111100} + \ket{11000011}) $\\
    $\ket{111}$ & $\frac{1}{\sqrt{2}}(\ket{10010110} + \ket{01101001})$ 
  \end{tabular}
  \caption{The computational basis as codewords}
  \label{tab:basis}
\end{table}

So, why is this code \emph{interesting}?  It is the smallest
member of the class of \emph{quasi-transversal} codes
\cite{Campbell2016Unifying-gate-s,PhysRevA.95.022316}, which allow the
efficient synthesis of multiqubit non-Clifford gates from noisy magic states
\cite{Sergei-Bravyi:2005eu}. 
For our purposes, its key property is that the logical
3-qubit doubly-controlled $Z$ (CCZ) 
gate is transversal, and can be implemented using the
single-qubit non-Clifford $T$ gate.
\[
CCZ = \mathrm{diag}(1,1,\ldots,1,-1)
\qquad\qquad
T = \begin{pmatrix}
1 & 0 \\
0 & e^{i\pi/4}
\end{pmatrix}\,
\]
More precisely, the CCZ is obtained by applying $T$ to black vertices
and \(T^\dagger\) to white vertices.  Therefore this code enables the
distillation of 8 noisy T-states into a less noisy CCZ logical
state. Since CCZ is Clifford equivalent to a Toffoli gate, this code
provides a basis for fault-tolerant quantum computation.  Therefore it
is a \emph{particularly} interesting test case for the use of
\zxcalculus and automated reasoning.

\section{The \zxcalculus}
\label{sec:zxcalculus}

The \zxcalculus \cite{Coecke:2009aa} is a universal formal graphical
notation for representing quantum states and processes, and a
sound\footnote{By \emph{sound} we mean every equation derivable in
  the \zxcalculus calculus is true in the standard Hilbert space
  interpretation; by \emph{complete} we mean that every true statement
in the Hilbert space model is derivable in the \zxcalculus.  Note that
the calculus is \emph{not} complete in this sense, but is complete
for \emph{fragments} of quantum theory.}
equational theory for reasoning about them.  In this section we
briefly review its syntax and semantics.

Since its introduction in \cite{Coecke:2008nx}, the \zxcalculus has
undergone various refinements, and various changes to the axioms have
been considered; for examples see
\cite{Duncan:2009ph,EPTCS171.5,Backens2016A-Simplified-St,Perdrix:2015aa}.
In this paper, we use the \emph{unconditional} calculus,
\emph{without} scalars\footnote{That is, for simplicity we will assume
  all the scalars are 1 and drop them whenever they occur; this is
  harmless, since in this work the zero scalar will not arise, and all
  the rest can be restored if needed.  See Backens
  \cite{Backens:2015aa} for a rigorous treatment of scalar factors in
  the \zxcalculus.}, \emph{without} supplementarity, and we take the
Hadamard gate as a \emph{primitive} element.  This version of the
calculus is complete for the stabilizer fragment of quantum mechanics
\cite{1367-2630-16-9-093021}, and for the 1-qubit Clifford+$T$ group
\cite{Backens:2014aa}; however it is not complete for the full,
multi-qubit, Clifford+$T$ group \cite{Perdrix:2015aa}, a fact which
will be pertinent later.

\begin{figure}
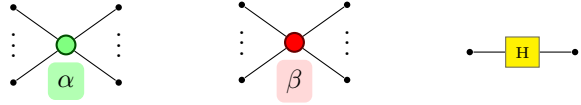

  \centering
  \[
\input{figures/zx-definition/Z.tikz}
\qquad \qquad
\input{figures/zx-definition/X.tikz}
\qquad \qquad
\input{figures/zx-definition/H.tikz} 
\]
  \caption{Allowed interior vertices}
  \label{fig:vertices}
\end{figure}

\begin{definition}
  \label{def:zxterms}
  A \emph{term} of the \zxcalculus is a finite open graph whose boundary is
  partitioned into \emph{inputs} and \emph{outputs}, and whose
  interior vertices are of the following types:
  \begin{itemize}
  \item $Z(\alpha)$ vertices, labelled by an angle $\alpha$, $0 \leq \alpha <
    2\pi$.  These are depicted as green or light grey circles; if $\alpha = 0$
    then the label is omitted.
  \item $X(\beta)$ vertices, labelled by an angle $\beta$, $0 \leq \beta <
    2\pi$.  These are are depicted as red or dark grey circles; again,
    if $\beta = 0$ then the label is omitted.
  \item $H$ vertices; unlike the other types H vertices are constrained
    to have degree exactly 2.  They are depicted as yellow squares.  
  \end{itemize}
  The allowed vertex types are shown in Figure \ref{fig:vertices}.  We
  adopt the convention that inputs are on the left, and outputs on the
  right.
\end{definition}

Terms, also called \emph{diagrams}, may be composed by joining some
number (maybe zero) of the outputs of one term to the inputs of
another.  Given a diagram $D : n \to m$ we define its \emph{adjoint}
$D^\dag : m \to n$ to be the diagram obtained by reflecting the
diagram around the vertical axis and negating all the angles.  The
terms of the \zxcalculus naturally form a $\dag$-symmetric monoidal
category \cite{Coecke:2015aa}, or more precisely a $\dag$-PROP
\cite{Lack:2004sf,Duncan2016Interacting-Fro}, which we call
\textbf{ZX}.  The only use we will make of this fact is defining the
semantics as a (strict) monoidal functor.  Note that the diagrams with
a single interior vertex generate all the rest by composition.

\begin{definition}\label{def:zxsemantics}
  Given a \textsl{zx}-term $D:n\to m$, its \emph{interpretation} is a
  linear map $\denote{D} : (\mathbb{C}^2)^{\otimes n} \to
  \mathbb{C}^2)^{\otimes m}$ defined as follows:
\[
    \ldenote{\input{figures/zx-definition/Z.tikz}}
     = 
    \left\{
    \begin{array}{ccl}
      \ket{0}^{\otimes m} & \mapsto &   \ket{0}^{\otimes n}\\
      \ket{1}^{\otimes m} & \mapsto &   e^{i \alpha}\ket{1}^{\otimes n}
    \end{array}\right.
\qquad\qquad
    \ldenote{\input{figures/zx-definition/X.tikz}}
     = 
    \left\{
    \begin{array}{ccl}
      \ket{+}^{\otimes m} & \mapsto &   \ket{+}^{\otimes n}\\
      \ket{-}^{\otimes m} & \mapsto &   e^{i \beta}\ket{-}^{\otimes n}
    \end{array}\right.
\]
\[
    \ldenote{\input{figures/zx-definition/H.tikz}}
    = 
    \frac{1}{\sqrt{2}}
    \left(
      \begin{array}{cc}
        1&1\\1&-1
      \end{array}
    \right).
\]
Since the above diagrams are the generators of \textbf{ZX}, the above
definition extends uniquely to a strict $\dag$-symmetric monoidal
functor $\denote{\cdot} : \mathbf{ZX} \to \fdhilb$.
\end{definition}

Any quantum circuit can be translated into the \zxcalculus via the
universal gate set shown.
  \[
  \begin{array}{ccccc}
    \circX{\alpha}
    & = &    \begin{pmatrix*}[r]
      \cos \frac{\alpha}{2} & -i\sin\frac{\alpha}{2} \\
      -i\sin\frac{\alpha}{2} & \cos \frac{\alpha}{2}  
    \end{pmatrix*}
    &=& \ldenote{\input{figures/zx-definition/X_alpha.tikz}} 
    \\ \\
    \circZ{\beta}
    &= &     \begin{pmatrix*}[r]
      1 & 0 \\ 0 & e^{i\beta}
    \end{pmatrix*}
    &= & \ldenote{\input{figures/zx-definition/Z_beta.tikz}}
    \\ \\
    \circH
    &= & \frac{1}{\sqrt{2}}
    \begin{pmatrix*}[r]
      1 & 1 \\ 1 & -1
    \end{pmatrix*}
    & = & \ldenote{\input{figures/zx-definition/H.tikz}}
    \\ \\
    \circCX
    &= &     \begin{pmatrix*}
      1 & 0 & 0 & 0\\
      0 & 1 & 0 & 0\\
      0 & 0 & 0 & 1\\
      0 & 0 & 1 & 0
    \end{pmatrix*}
    &= & \ldenote{\input{figures/zx-definition/CNOT.tikz}}
  \end{array}
  \]
  Basic state preparations have a simple form:
  \[
\ket{0} =
\ldenote{\redunit}
\qquad\qquad
\ket{1} =
\ldenote{\redpoint{\pi}}
\qquad\qquad
\ket{+} =
\ldenote{\greenunit}
\qquad\qquad
\ket{-} =
\ldenote{\greenpoint{\pi}}
  \]
However not all gates have simple translations.  Using
the standard decomposition \cite{Adriano-Barenco:1995qy}, 
the CCZ is translated via CNOTs and single-qubit phase gates, leading
to quite a complex diagram.
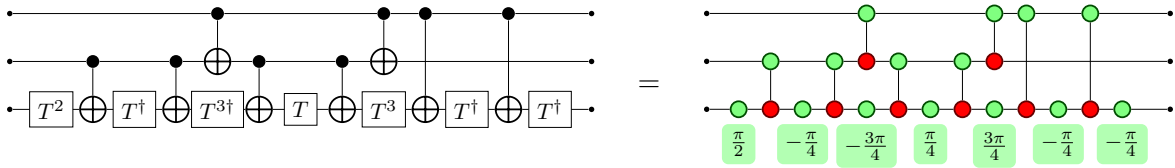
\[
\scalebox{0.85}{%
\beginpgfgraphicnamed{circuit-ccz}
\InputIfFileExists{circuit-ccz.tikz}{}{\input{./figures/circuit-ccz.tikz}}
\endpgfgraphicnamed} \quad = \quad
\scalebox{0.85}{%
\beginpgfgraphicnamed{ccz-circuit}
\InputIfFileExists{ccz-circuit.tikz}{}{\input{./figures/ccz-circuit.tikz}}
\endpgfgraphicnamed}
\]  
This diagram can be reduced to a simpler form, but not greatly so.
Note the use of $T = Z(\frac{\pi}{4})$ gates makes explicit that this
diagram is outside the Clifford fragment and in the realm where the
\zxcalculus is not complete.

The \zxcalculus has a rich equational theory based on the theory of
Frobenius-Hopf algebras
\cite{Coecke:2009aa,Duncan2016Interacting-Fro}.  A
concise\footnote{Though not minimal; see
  \cite{Backens2016A-Simplified-St}.} presentation of the axioms is
shown in Figure \ref{fig:zxrules}.  As noted earlier, the equations
are sound\footnote{For each equation
  $L = R$ we have 
  $\denote{L} = z\denote{R}$ for some non-zero $z\in\mathbb C$.
  Soundness on the nose can be achieved by explicitly representing the
  scalars, but for expository purposes we prefer to keep the diagrams
  simple.} with respect to $\denote{\cdot}$.

\begin{figure}[t]

  \makebox[\textwidth][c]{
  \begin{minipage}{1.1\linewidth}
    \[
    \begin{array}{ccccc}
      \input{figures/zx-definition/axioms/spider_1.tikz}
      \;\;=\;\;
      \input{figures/zx-definition/axioms/spider_2.tikz}
      & 
      \qquad 
      &
      \input{figures/zx-definition/axioms/anti-loop_1.tikz}
      \;\;=\;\;
      \input{figures/zx-definition/axioms/anti-loop_2.tikz}
      &
      \qquad 
      &
      \input{figures/zx-definition/axioms/identity_1.tikz}
      \;\;=\;\;
      \input{figures/zx-definition/axioms/identity_2.tikz}
      \\
      \text{(spider)} && \text{(anti-loop)} && \text{(identity)} 
      \\ \\
      \input{figures/zx-definition/axioms/copying_1.tikz}
      \;\;=\;\;
      \input{figures/zx-definition/axioms/copying_2.tikz}
      &
      \qquad 
      &
      \input{figures/zx-definition/axioms/pi-commute_1.tikz}
      \;\;=\;\;
      \input{figures/zx-definition/axioms/pi-commute_2.tikz}
      &
      \qquad 
      &
      \input{figures/zx-definition/axioms/hopf_1.tikz}
      \;\;=\;\;
      \input{figures/zx-definition/axioms/hopf_2.tikz}
      \\ 
      \text{(copying)}  &&  \text{($\pi$-commute)} &&\text{(hopf)} 
      \\ \\
      \input{figures/zx-definition/axioms/bialgebra_1.tikz}
      \;\;=\;\;
      \input{figures/zx-definition/axioms/bialgebra_2.tikz}
      & 
      \qquad
      & 
      \input{figures/zx-definition/axioms/h-commute_1.tikz}
      \;\;=\;\;
      \input{figures/zx-definition/axioms/h-commute_2.tikz}
      & 
      \qquad 
      &
      \input{figures/zx-definition/H.tikz}
      \;\;=\;\;
      \input{figures/zx-definition/axioms/h-euler-2.tikz}
      \\ \\[-1ex]
      \text{(bialgebra)}  &&\text{($H$-commute)}  &&\text{($H$-euler)} 
    \end{array}
    \]
  \end{minipage}}
  \caption{The rules of the \zxcalculus}
  \label{fig:zxrules}
  \figureline
\end{figure}

\section{Working with Quantomatic}
\label{sec:quantomatic}

We now give a brief overview of Quantomatic for the uninitiated.  For
a full description of the system see
\cite{Kissinger2015Quantomatic:-A-}; to obtain it, see
\cite{quantomatichome}.

Quantomatic is an interactive theorem prover 
which can prove equations between terms of 
the \zxcalculus.  
The user draws the desired term in the
graphical editor, and builds the proof by applying \emph{rewrite rules} to
the current graph.  A rewrite rule is a \emph{directed} equation; \ie
two graphs which have the same boundary.  The user selects the rule or
rules they wish to apply and Quantomatic will display 
where the LHS of the rule matches a subgraph of the current term,
alongside the term obtained by replacing the matched subgraph with the
RHS of the rule.  A proof consists of a sequence of terms linked
by the application of a particular rule at a particular location in
the term.

A Quantomatic proof development begins with a set of axioms,  \ie
rules without proofs which are justified semantically.  Whereas in
pen-and-paper presentations (like Figure \ref{fig:zxrules}) we might
aim for a minimal set of rules, for automated proof developments a
larger, possibly redundant, set of rules is usually more
convenient\footnote{The axiom ruleset used for this project is given
  in Appendix \ref{sec:ruleset}.}.  However 
any proof can be promoted to a lemma, which can then be
used as a rule.  In the course of a proof development the user will
typically build up a collection of increasingly powerful lemmas
encapsulating pieces of local reasoning within the larger proof.  In
some cases the lemmas may simply be special cases of more general
axioms intended to cut down the search space of possible rule
applications.

Quantomatic also supports automated deduction via simplification
procedures, aka \emph{simprocs}.  A simproc is a simple program built
from combinators which allow the sequencing of rules, greedy
application of every rule in a given ruleset, reduction according to
some rules until some condition is met, and so on. 
Simprocs allow quite lengthy
proofs to be carried out automatically.
However some caution is required.
The complete \zxcalculus is neither terminating nor
confluent, so the rules to be included in a simproc must be chosen
with care.  Termination can be ensured by using only rules where the LHS is
in some sense strictly ``bigger'' than the RHS; obviously the graph cannot get
smaller forever.  However many essential
rules (\eg $\pi$-commute) do not fit this pattern.
The other danger is that
by using a simproc we give up control of \emph{where} in the term
rewrites will be applied; lack of confluence means it is possible for
the simproc to make a bad choice, from which it cannot reach the
desired final term.  In practice this means that a proof often
consists of an alternation of simproc application and human selected
rewrites which expand the graph, although depending on the proof it is
sometimes possible to formalise this alternation itself in a simproc.

\begin{figure}
  \centering
\begin{verbatim}
val red_pi_lemma = load_rule "created_theorms/pauli/red_pi_lemma";
val red_sp = load_rule "axioms/red_sp";

val reduction_before_pi = load_ruleset [
  "axioms/red_copy", "axioms/red_sp", "axioms/green_sp", "axioms/hopf",
  "axioms/red_scalar", "axioms/green_scalar", "axioms/green_id",
  "axioms/red_id", "axioms/red_loop", "axioms/green_loop"
];

val simproc = (
  LOOP (
    REDUCE_ALL reduction_before_pi ++
    REWRITE red_pi_lemma ++
    REDUCE red_sp
  )
);
\end{verbatim}
  \caption{An example simproc: \texttt{push\_pauli\_x}}
  \label{fig:pushpaulix}
\end{figure}

Some of the rules shown in Figure \ref{fig:zxrules} 
have ellipses, to indicate that the given diagram indicates an
infinite family of terms.  Such families are formalised in Quantomatic
using \emph{!-boxes}.  Briefly\footnote{Less briefly, see
  \cite{Aleks-Kissinger:2012qy,Kissinger:2014ab,Kissinger:2015aa,Quick2015-Logic}
  for a full treatment of !-boxes and their associated logic.}, a
!-box is a marked subgraph within a term, which indicates that the
subgraph may occur any number of times when matching the graph.  For
example, the $\pi$-commute (below left) rule may be formalised using
!-boxes as shown on the right,
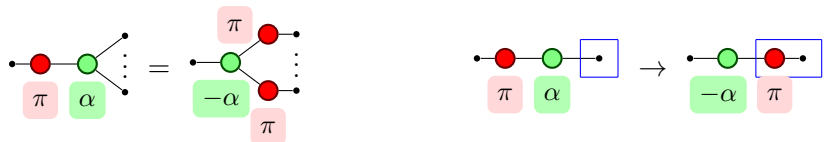
\[
      \input{figures/zx-definition/axioms/pi-commute_1.tikz}
      \;\;=\;\;
      \input{figures/zx-definition/axioms/pi-commute_2.tikz}
\qquad
\qquad
\qquad
      \input{figures/zx-definition/axioms/pi-commute_1_box.tikz}
      \;\; \to \;\;
      \input{figures/zx-definition/axioms/pi-commute_2_box.tikz}
\]
where the blue boxes are the indicated subgraphs.  Note that the
matching of the LHS in the target graph determines how many times the
box should be repeated, on both sides of the rule.  Many of the
multi-ary rules of the \zxcalculus are formalised using !-boxes; in
order to better control of the matching algorithm, it is often
necessary to derive fixed arity special cases as lemmas.


\section{The Encoder and Decoder}
\label{sec:encoder-decoder}

No circuit for encoding the logical qubits into physical qubits is
given in any of the references
\cite{Campbell:colourcode,Campbell2016Unifying-gate-s,PhysRevA.95.022316,PhysRevA.95.032338}
so our first task is to define the encoding operation in the
\zxcalculus.  Looking at Table \ref{tab:basis}, we can see that the
basis states are prepared by conditionally applying Pauli $X$
operations an 8-qubit GHZ state.  In the \zxcalculus GHZ states have a
very simple form:
\[
\input{figures/encoder/8-qubit-space.tikz}
= \quad \ket{00000000} + \ket{11111111}
\]
The bit flips are applied by ``copying'' a Pauli $X$ from a (logical)
input qubit to the appropriate four physical qubits, as shown below.
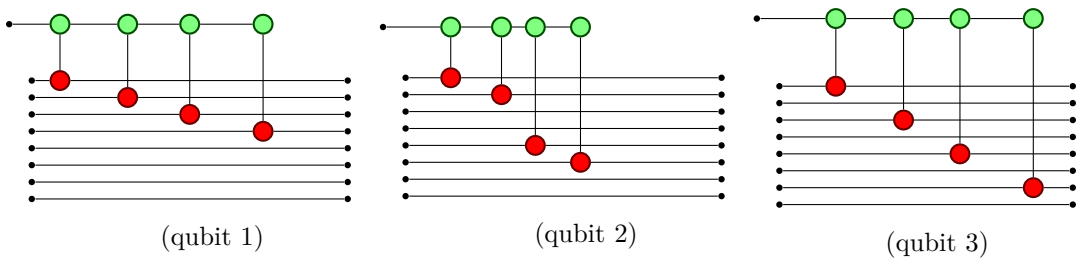
\[
\input{figures/encoder/enc-q1.tikz}
\quad
\input{figures/encoder/enc-q2.tikz}
\quad
\input{figures/encoder/enc-q3.tikz}
\]
The final encoder is obtained by composing these four diagrams along
the 8 physical qubits;  logical qubits come in on the left, codewords
come out on the right.  The decoder is simply the encoder in reverse.
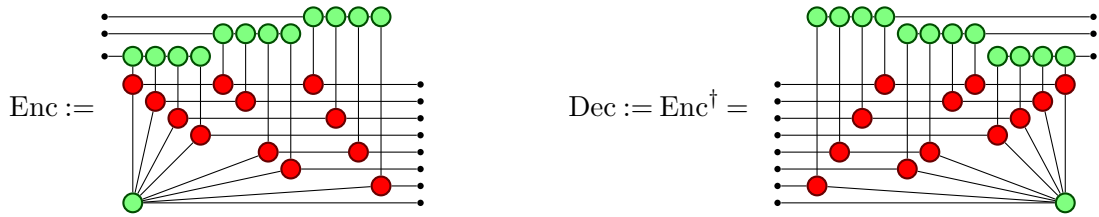
\[
\Enc : = 
\input{figures/encoder/encoder.tikz}
\quad 
\qquad \qquad
\Dec := \Enc^\dag = \ \ \input{figures/encoder/decoder.tikz}
\]
We note that the graph for \Enc contains no $H$ vertices and no
angles.  For such terms we have the following result:

\begin{proposition}{\cite{Kissinger2014Notes-on-Graphi}}
  \label{prop:rotatesimp}
  Any \zxcalculus term without angles or $H$ vertices is equal to a
  term in \emph{subset-spanning form}; further, the subset-spanning
  form is quasi-unique and can be computed by a terminating algorithm.
\end{proposition}

Space does not permit a description of the quasi-normal forms; the
important point here is that the normalisation procedure is
implemented in a Quantomatic simproc called \texttt{rotate\_simp}.  We
will make extensive use of this in the sequel.

\begin{proposition}
  \label{prop:encdec}
  Encoding followed by decoding is the identity, \ie $\Enc^\dag \circ \Enc
  = \id{3}$, or in pictures:
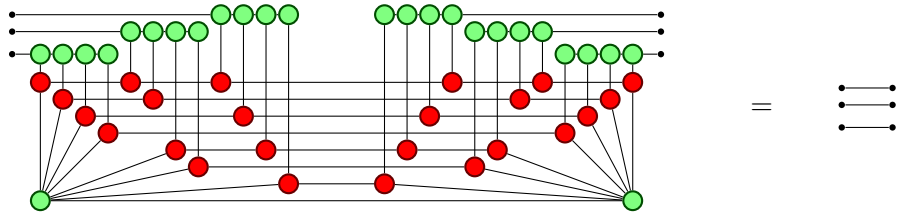
\[
\input{figures/encoder/enc-dec.tikz}
\qquad
=
\qquad
\input{figures/encoder/enc-dec_1.tikz}
\]
\end{proposition}
This result can be proven by \texttt{rotate\_simp} without human
intervention; it requires 66 rewrite steps and takes approximately 5
seconds of real time.  See Supplementary Material \ref{SUPP-sec:enc-identity}.  Note that
Proposition \ref{prop:encdec} doesn't establish that the 8 codewords
of Table \ref{tab:basis} are prepared as required; this is a corollary
of the tranversality of the Pauli group, which we show in the next
section.

The astute reader will have noticed that the encoder is not actually a
quantum circuit: the top ``rails'' end in a projection onto the state
$\ket+$.  However, no post-selection is required: we can implement
\Enc with a circuit using five ancilla qubits.

\begin{proposition}
  \label{prop:enccirc}
  \Enc is a unitary embedding.
  \begin{proof}
    We do the proof in two stages, shown below.
    \[
    \input{figures/encoder/encoder.tikz} 
    \quad = \quad
    \input{figures/encoder/encoder-NF.tikz} 
    \quad = \quad
    \input{figures/encoder/encoder-circuit.tikz}
    \]
    (For the full proof see Supplementary Material
    \ref{SUPP-sec:enc-circuit}.)  From the final diagram, one can
    easily read the translation back into conventional circuit
    notation.
    \[
    \scalebox{0.75}{\input{figures/encoder/encoder-circuit-1a.tikz}}
    \;\;\equiv \;\;
    \scalebox{0.75}{\input{figures/encoder/encoder-circuit-2.tikz}}
    \]
The first equation above is the normalisation of \Enc using
\texttt{rotate\_simp}; it takes 28 rewrite steps and 2 seconds.  The
second equation required substantial human intervention.  The proof
requires \emph{expanding} the graph rather than reducing it, orienting
the rewrite rules in the opposite direction to normal.  If naively
applied this would lead to an uncontrolled blow-up of the graph.
Further, even under human control, the normal ruleset produces too many
matches to be helpful.  For these reasons it was necessary to prove
some (near trivial) intermediate lemmas:
\begin{lem}
\begin{quote}\raggedright
\begin{tikzpicture}[baseline={([yshift=-.5ex]current bounding box.center)}, x=0.3cm, y=0.3cm]
	\begin{pgfonlayer}{nodelayer}
		\node [style=wire] (b1) at (-10.0, 2.25) {};
		\node [style=wire] (b0) at (-10.0, 4.0) {};
		\node [style=Z] (v3) at (-8.25, 2.25) {};
		\node [style=wire] (b2) at (-2.75, 4.0) {};
		\node [style=X] (v1) at (-4.75, 2.25) {};
		\node [style=X] (v0) at (-8.25, 4.0) {};
		\node [style=Z] (v2) at (-4.75, 4.0) {};
		\node [style=wire] (b3) at (-2.75, 2.25) {};
		\node (bx3ul) at (-10.4, 2.6333333333333333) {};
		\node [style=none] (bx3ur) at (-9.6, 2.6333333333333333) {};
		\node [style=none] (bx3ll) at (-10.4, 1.8666666666666667) {};
		\node [style=none] (bx3lr) at (-9.6, 1.8666666666666667) {};
		\node (bx1ul) at (-3.1333333333333333, 4.4) {};
		\node [style=none] (bx1ur) at (-2.3666666666666667, 4.4) {};
		\node [style=none] (bx1ll) at (-3.1333333333333333, 3.6) {};
		\node [style=none] (bx1lr) at (-2.3666666666666667, 3.6) {};
		\node (bx2ul) at (-3.1333333333333333, 2.6333333333333333) {};
		\node [style=none] (bx2ur) at (-2.3666666666666667, 2.6333333333333333) {};
		\node [style=none] (bx2ll) at (-3.1333333333333333, 1.8666666666666667) {};
		\node [style=none] (bx2lr) at (-2.3666666666666667, 1.8666666666666667) {};
		\node (bx0ul) at (-10.4, 4.4) {};
		\node [style=none] (bx0ur) at (-9.6, 4.4) {};
		\node [style=none] (bx0ll) at (-10.4, 3.6) {};
		\node [style=none] (bx0lr) at (-9.6, 3.6) {};
		\node [style=none] (padr) at (-1.3666666666666667, 5.4) {};
		\node [style=none] (padd) at (-1.3666666666666667, 0.8666666666666667) {};
	\end{pgfonlayer}
	\begin{pgfonlayer}{edgelayer}
		\draw [style=simple] (v1) to (b3);
		\draw [style=simple] (v2) to (v0);
		\draw [style=simple] (v0) to (v3);
		\draw [style=simple] (v2) to (v1);
		\draw [style=simple] (v3) to (v1);
		\draw [style=simple] (b2) to (v2);
		\draw [style=simple] (b0) to (v0);
		\draw [style=simple] (v3) to (b1);
		\draw [style=blue] (bx3ul.center) to (bx3ur.center);
		\draw [style=blue] (bx3ul.center) to (bx3ll.center);
		\draw [style=blue] (bx3ll.center) to (bx3lr.center);
		\draw [style=blue] (bx3lr.center) to (bx3ur.center);
		\draw [style=blue] (bx1ul.center) to (bx1ur.center);
		\draw [style=blue] (bx1ul.center) to (bx1ll.center);
		\draw [style=blue] (bx1ll.center) to (bx1lr.center);
		\draw [style=blue] (bx1lr.center) to (bx1ur.center);
		\draw [style=blue] (bx2ul.center) to (bx2ur.center);
		\draw [style=blue] (bx2ul.center) to (bx2ll.center);
		\draw [style=blue] (bx2ll.center) to (bx2lr.center);
		\draw [style=blue] (bx2lr.center) to (bx2ur.center);
		\draw [style=blue] (bx0ul.center) to (bx0ur.center);
		\draw [style=blue] (bx0ul.center) to (bx0ll.center);
		\draw [style=blue] (bx0ll.center) to (bx0lr.center);
		\draw [style=blue] (bx0lr.center) to (bx0ur.center);
	\end{pgfonlayer}
\end{tikzpicture}
=
\begin{tikzpicture}[baseline={([yshift=-.5ex]current bounding box.center)}, x=0.3cm, y=0.3cm]
	\begin{pgfonlayer}{nodelayer}
		\node [style=Z] (v4) at (-6.5, 4.0) {};
		\node [style=wire] (b1) at (-10.0, 2.25) {};
		\node [style=wire] (b0) at (-10.0, 4.0) {};
		\node [style=X] (v1bx1) at (-8.25, 4.0) {};
		\node [style=Z] (v0bx3) at (-4.5, 4.0) {};
		\node [style=wire] (b2) at (-2.75, 4.0) {};
		\node [style=X] (v1bx0) at (-4.25, 2.25) {};
		\node [style=X] (v5) at (-6.5, 2.25) {};
		\node [style=wire] (b3) at (-2.75, 2.25) {};
		\node [style=Z] (v0bx2) at (-8.25, 2.25) {};
		\node (bx3ul) at (-10.4, 2.6333333333333333) {};
		\node [style=none] (bx3ur) at (-9.6, 2.6333333333333333) {};
		\node [style=none] (bx3ll) at (-10.4, 1.8666666666666667) {};
		\node [style=none] (bx3lr) at (-9.6, 1.8666666666666667) {};
		\node (bx1ul) at (-3.1333333333333333, 4.4) {};
		\node [style=none] (bx1ur) at (-2.3666666666666667, 4.4) {};
		\node [style=none] (bx1ll) at (-3.1333333333333333, 3.6) {};
		\node [style=none] (bx1lr) at (-2.3666666666666667, 3.6) {};
		\node (bx2ul) at (-3.1333333333333333, 2.6333333333333333) {};
		\node [style=none] (bx2ur) at (-2.3666666666666667, 2.6333333333333333) {};
		\node [style=none] (bx2ll) at (-3.1333333333333333, 1.8666666666666667) {};
		\node [style=none] (bx2lr) at (-2.3666666666666667, 1.8666666666666667) {};
		\node (bx0ul) at (-10.4, 4.4) {};
		\node [style=none] (bx0ur) at (-9.6, 4.4) {};
		\node [style=none] (bx0ll) at (-10.4, 3.6) {};
		\node [style=none] (bx0lr) at (-9.6, 3.6) {};
		\node [style=none] (padl) at (-11.4, 0.8666666666666667) {};
		\node [style=none] (padr) at (-1.3666666666666667, 5.4) {};
		\node [style=none] (padu) at (-11.4, 5.4) {};
		\node [style=none] (padd) at (-1.3666666666666667, 0.8666666666666667) {};
	\end{pgfonlayer}
	\begin{pgfonlayer}{edgelayer}
		\draw [style=simple] (b3) to (v1bx0);
		\draw [style=simple] (v5) to (v0bx2);
		\draw [style=simple] (v5) to (v0bx3);
		\draw [style=simple] (v0bx3) to (b2);
		\draw [style=simple] (b0) to (v1bx1);
		\draw [style=simple] (v0bx2) to (b1);
		\draw [style=simple] (v4) to (v5);
		\draw [style=simple] (v1bx0) to (v4);
		\draw [style=simple] (v1bx1) to (v4);
		\draw [style=blue] (bx3ul.center) to (bx3ur.center);
		\draw [style=blue] (bx3ul.center) to (bx3ll.center);
		\draw [style=blue] (bx3ll.center) to (bx3lr.center);
		\draw [style=blue] (bx3lr.center) to (bx3ur.center);
		\draw [style=blue] (bx1ul.center) to (bx1ur.center);
		\draw [style=blue] (bx1ul.center) to (bx1ll.center);
		\draw [style=blue] (bx1ll.center) to (bx1lr.center);
		\draw [style=blue] (bx1lr.center) to (bx1ur.center);
		\draw [style=blue] (bx2ul.center) to (bx2ur.center);
		\draw [style=blue] (bx2ul.center) to (bx2ll.center);
		\draw [style=blue] (bx2ll.center) to (bx2lr.center);
		\draw [style=blue] (bx2lr.center) to (bx2ur.center);
		\draw [style=blue] (bx0ul.center) to (bx0ur.center);
		\draw [style=blue] (bx0ul.center) to (bx0ll.center);
		\draw [style=blue] (bx0ll.center) to (bx0lr.center);
		\draw [style=blue] (bx0lr.center) to (bx0ur.center);
	\end{pgfonlayer}
\end{tikzpicture}
\small{(bialgebra x1)}
\end{quote}

\end{lem}
\begin{lem}
\begin{quote}\raggedright
\begin{tikzpicture}[baseline={([yshift=-.5ex]current bounding box.center)}, x=0.3cm, y=0.3cm]
	\begin{pgfonlayer}{nodelayer}
		\node [style=wire] (b1) at (-10.25, 2.0) {};
		\node [style=wire] (b0) at (-10.25, 3.75) {};
		\node [style=Z] (v3) at (-4.25, 3.75) {};
		\node [style=wire] (b2) at (-1.25, 3.75) {};
		\node [style=X] (v1) at (-4.25, 2.0) {};
		\node [style=X] (v0) at (-7.25, 3.75) {};
		\node [style=Z] (v2) at (-7.25, 2.0) {};
		\node [style=wire] (b3) at (-1.5, 2.0) {};
		\node (bx1ul) at (-1.9, 2.4) {};
		\node [style=none] (bx1ur) at (-1.1, 2.4) {};
		\node [style=none] (bx1ll) at (-1.9, 1.6) {};
		\node [style=none] (bx1lr) at (-1.1, 1.6) {};
		\node (bx0ul) at (-10.633333333333333, 4.133333333333334) {};
		\node [style=none] (bx0ur) at (-9.866666666666667, 4.133333333333334) {};
		\node [style=none] (bx0ll) at (-10.633333333333333, 3.3666666666666667) {};
		\node [style=none] (bx0lr) at (-9.866666666666667, 3.3666666666666667) {};
		\node [style=none] (padr) at (-0.10000000000000009, 5.133333333333334) {};
		\node [style=none] (padd) at (-0.10000000000000009, 0.6000000000000001) {};
	\end{pgfonlayer}
	\begin{pgfonlayer}{edgelayer}
		\draw [style=simple] (v1) to (b3);
		\draw [style=simple] (v0) to (v2);
		\draw [style=simple] (v3) to (v0);
		\draw [style=simple] (v2) to (v1);
		\draw [style=simple] (v3) to (v1);
		\draw [style=simple] (b2) to (v3);
		\draw [style=simple] (b0) to (v0);
		\draw [style=simple] (v2) to (b1);
		\draw [style=blue] (bx1ul.center) to (bx1ur.center);
		\draw [style=blue] (bx1ul.center) to (bx1ll.center);
		\draw [style=blue] (bx1ll.center) to (bx1lr.center);
		\draw [style=blue] (bx1lr.center) to (bx1ur.center);
		\draw [style=blue] (bx0ul.center) to (bx0ur.center);
		\draw [style=blue] (bx0ul.center) to (bx0ll.center);
		\draw [style=blue] (bx0ll.center) to (bx0lr.center);
		\draw [style=blue] (bx0lr.center) to (bx0ur.center);
	\end{pgfonlayer}
\end{tikzpicture}
=
\begin{tikzpicture}[baseline={([yshift=-.5ex]current bounding box.center)}, x=0.3cm, y=0.3cm]
	\begin{pgfonlayer}{nodelayer}
		\node [style=Z] (v4) at (-4.75, 3.5) {};
		\node [style=wire] (b1) at (-8.5, 1.5) {};
		\node [style=wire] (b0) at (-8.5, 3.5) {};
		\node [style=X] (v1bx1) at (-6.75, 3.5) {};
		\node [style=wire] (b2) at (-0.5, 3.75) {};
		\node [style=X] (v1bx0) at (-2.75, 1.5) {};
		\node [style=X] (v5) at (-4.75, 1.5) {};
		\node [style=wire] (b3) at (-0.5, 1.5) {};
		\node (bx1ul) at (-0.9, 1.9) {};
		\node [style=none] (bx1ur) at (-0.1, 1.9) {};
		\node [style=none] (bx1ll) at (-0.9, 1.1) {};
		\node [style=none] (bx1lr) at (-0.1, 1.1) {};
		\node (bx0ul) at (-8.9, 3.9) {};
		\node [style=none] (bx0ur) at (-8.1, 3.9) {};
		\node [style=none] (bx0ll) at (-8.9, 3.1) {};
		\node [style=none] (bx0lr) at (-8.1, 3.1) {};
		\node [style=none] (padl) at (-9.9, 0.10000000000000009) {};
		\node [style=none] (padr) at (0.9, 4.9) {};
		\node [style=none] (padu) at (-9.9, 4.9) {};
		\node [style=none] (padd) at (0.9, 0.10000000000000009) {};
	\end{pgfonlayer}
	\begin{pgfonlayer}{edgelayer}
		\draw [style=simple] (b3) to (v1bx0);
		\draw [style=simple] (v5) to (b2);
		\draw [style=simple] (b0) to (v1bx1);
		\draw [style=simple] (v5) to (b1);
		\draw [style=simple] (v4) to (v5);
		\draw [style=simple] (v1bx0) to (v4);
		\draw [style=simple] (v1bx1) to (v4);
		\draw [style=blue] (bx1ul.center) to (bx1ur.center);
		\draw [style=blue] (bx1ul.center) to (bx1ll.center);
		\draw [style=blue] (bx1ll.center) to (bx1lr.center);
		\draw [style=blue] (bx1lr.center) to (bx1ur.center);
		\draw [style=blue] (bx0ul.center) to (bx0ur.center);
		\draw [style=blue] (bx0ul.center) to (bx0ll.center);
		\draw [style=blue] (bx0ll.center) to (bx0lr.center);
		\draw [style=blue] (bx0lr.center) to (bx0ur.center);
	\end{pgfonlayer}
\end{tikzpicture}
\small{(bialgebra x1, Z identity x2)}
\end{quote}

\end{lem}
\begin{lem}
\begin{quote}\raggedright
\begin{tikzpicture}[baseline={([yshift=-.5ex]current bounding box.center)}, x=0.3cm, y=0.3cm]
	\begin{pgfonlayer}{nodelayer}
		\node [style=wire] (b1) at (-13.0, 2.0) {};
		\node [style=wire] (b0) at (-13.0, 4.25) {};
		\node [style=Z] (v3) at (-10.0, 2.0) {};
		\node [style=wire] (b2) at (-4.25, 4.25) {};
		\node [style=X] (v1) at (-8.0, 2.0) {};
		\node [style=X] (v0) at (-10.0, 4.25) {};
		\node [style=Z] (v2) at (-8.0, 4.25) {};
		\node [style=wire] (b3) at (-4.25, 2.0) {};
		\node (bx1ul) at (-13.4, 2.3833333333333333) {};
		\node [style=none] (bx1ur) at (-12.6, 2.3833333333333333) {};
		\node [style=none] (bx1ll) at (-13.4, 1.6166666666666667) {};
		\node [style=none] (bx1lr) at (-12.6, 1.6166666666666667) {};
		\node (bx0ul) at (-4.633333333333334, 4.65) {};
		\node [style=none] (bx0ur) at (-3.8666666666666667, 4.65) {};
		\node [style=none] (bx0ll) at (-4.633333333333334, 3.85) {};
		\node [style=none] (bx0lr) at (-3.8666666666666667, 3.85) {};
		\node [style=none] (padr) at (-2.8666666666666667, 5.65) {};
		\node [style=none] (padd) at (-2.8666666666666667, 0.6166666666666667) {};
	\end{pgfonlayer}
	\begin{pgfonlayer}{edgelayer}
		\draw [style=simple] (v1) to (b3);
		\draw [style=simple] (v0) to (v2);
		\draw [style=simple] (v0) to (v3);
		\draw [style=simple] (v2) to (v1);
		\draw [style=simple] (v3) to (v1);
		\draw [style=simple] (v2) to (b2);
		\draw [style=simple] (b0) to (v0);
		\draw [style=simple] (b1) to (v3);
		\draw [style=blue] (bx1ul.center) to (bx1ur.center);
		\draw [style=blue] (bx1ul.center) to (bx1ll.center);
		\draw [style=blue] (bx1ll.center) to (bx1lr.center);
		\draw [style=blue] (bx1lr.center) to (bx1ur.center);
		\draw [style=blue] (bx0ul.center) to (bx0ur.center);
		\draw [style=blue] (bx0ul.center) to (bx0ll.center);
		\draw [style=blue] (bx0ll.center) to (bx0lr.center);
		\draw [style=blue] (bx0lr.center) to (bx0ur.center);
	\end{pgfonlayer}
\end{tikzpicture}
=
\begin{tikzpicture}[baseline={([yshift=-.5ex]current bounding box.center)}, x=0.3cm, y=0.3cm]
	\begin{pgfonlayer}{nodelayer}
		\node [style=Z] (v4) at (-8.5, 4.25) {};
		\node [style=wire] (b1) at (-13.0, 2.0) {};
		\node [style=wire] (b0) at (-13.0, 4.25) {};
		\node [style=Z] (v0bx3) at (-7.0, 2.0) {};
		\node [style=wire] (b2) at (-4.25, 4.25) {};
		\node [style=X] (v5) at (-8.5, 2.0) {};
		\node [style=wire] (b3) at (-4.25, 2.0) {};
		\node [style=Z] (v0bx2) at (-10.25, 2.0) {};
		\node (bx1ul) at (-13.4, 2.3833333333333333) {};
		\node [style=none] (bx1ur) at (-12.6, 2.3833333333333333) {};
		\node [style=none] (bx1ll) at (-13.4, 1.6166666666666667) {};
		\node [style=none] (bx1lr) at (-12.6, 1.6166666666666667) {};
		\node (bx0ul) at (-4.633333333333334, 4.65) {};
		\node [style=none] (bx0ur) at (-3.8666666666666667, 4.65) {};
		\node [style=none] (bx0ll) at (-4.633333333333334, 3.85) {};
		\node [style=none] (bx0lr) at (-3.8666666666666667, 3.85) {};
		\node [style=none] (padl) at (-14.4, 0.6166666666666667) {};
		\node [style=none] (padr) at (-2.8666666666666667, 5.65) {};
		\node [style=none] (padu) at (-14.4, 5.65) {};
		\node [style=none] (padd) at (-2.8666666666666667, 0.6166666666666667) {};
	\end{pgfonlayer}
	\begin{pgfonlayer}{edgelayer}
		\draw [style=simple] (b3) to (v4);
		\draw [style=simple] (v5) to (v0bx2);
		\draw [style=simple] (v5) to (v0bx3);
		\draw [style=simple] (v0bx3) to (b2);
		\draw [style=simple] (b0) to (v4);
		\draw [style=simple] (v0bx2) to (b1);
		\draw [style=simple] (v4) to (v5);
		\draw [style=blue] (bx1ul.center) to (bx1ur.center);
		\draw [style=blue] (bx1ul.center) to (bx1ll.center);
		\draw [style=blue] (bx1ll.center) to (bx1lr.center);
		\draw [style=blue] (bx1lr.center) to (bx1ur.center);
		\draw [style=blue] (bx0ul.center) to (bx0ur.center);
		\draw [style=blue] (bx0ul.center) to (bx0ll.center);
		\draw [style=blue] (bx0ll.center) to (bx0lr.center);
		\draw [style=blue] (bx0lr.center) to (bx0ur.center);
	\end{pgfonlayer}
\end{tikzpicture}
\small{(bialgebra x1, X identity x2)}
\end{quote}

\end{lem}
The sharp-eyed will notice that each of these lemmas introduces an
explicit CNOT into the circuit.  However, they are not mathematically
interesting statements: they are chosen to have few matches on the
in-progress proof term, and hence take control of the matching
algorithm.
  \end{proof}
\end{proposition}

\section{Fault-tolerant operations}
\label{sec:fault-toler-oper}

Given our error correcting scheme, we are interested in the
\emph{fault-tolerant operations}.  That is, unitary operators on the
codeword space which commute with the encoder.  More precisely, given
a unitary $L$ acting on the logical Hilbert space, and a unitary $P$
acting on the physical Hilbert space we say that $L$ is the fault
tolerant version of $P$ if the equation
\begin{equation}
\Enc \circ L = P \circ \Enc\label{eq:1}
\end{equation}
holds.  Since we don't really care what $P$ does to non-codewords, and
by virtue of Proposition \ref{prop:encdec}, the weaker equation 
\begin{equation}
L = \Enc^\dag \circ P \circ \Enc\label{eq:2}
\end{equation}
will suffice; given the nature of our tools, reducing a big
complicated graph to a small one will usually be easier than showing
that two rough-equally-complex graphs are equal.  Statements of the
form (\ref{eq:1}) are most easily proven by rewriting both sides of
the equation to a common reduct.

\subsection{The Paulis}
\label{sec:paulis}

As claimed in Section \ref{sec:code}, the Pauli gates are transversal
for this code.

\begin{proposition}\label{prop:paulis}
  We have the following equations:
  \begin{align}
    \Enc \circ (1 \otimes 1 \otimes X)  & = (X \otimes X \otimes X \otimes X \otimes
    1 \otimes 1 \otimes 1 \otimes 1) \circ \Enc \label{eq:x1} \tag{x1} \\
    \Enc \circ (1 \otimes X \otimes 1)  & = (X \otimes X \otimes 1 \otimes 1 \otimes
    X \otimes X \otimes 1 \otimes 1) \circ \Enc  \tag{x2}\label{eq:x2} \\
    \Enc \circ (X \otimes 1 \otimes 1)  & = (X \otimes 1 \otimes X \otimes 1 \otimes
    X \otimes 1 \otimes X \otimes 1) \circ \Enc  \tag{x3}\label{eq:x3} \\
    \Enc \circ (1 \otimes 1 \otimes Z)  & = (Z \otimes 1 \otimes 1 \otimes 1 \otimes
    Z \otimes 1 \otimes 1 \otimes 1) \circ \Enc \tag{z1}\label{eq:z1} \\
    \Enc \circ (1 \otimes Z \otimes 1)  & = (Z \otimes 1 \otimes Z \otimes 1 \otimes
    1 \otimes 1 \otimes 1 \otimes 1) \circ \Enc \tag{z2}\label{eq:z2}\\
    \Enc \circ (Z \otimes 1 \otimes 1)  & = (Z \otimes Z \otimes 1 \otimes 1 \otimes
    1 \otimes 1 \otimes 1 \otimes 1) \circ \Enc \tag{z3}\label{eq:z3}
  \end{align}
  \begin{proof}
    We demonstrate \eqref{eq:x1} by rewriting as shown below.
    \begin{center}
      \makebox[\textwidth][c]{
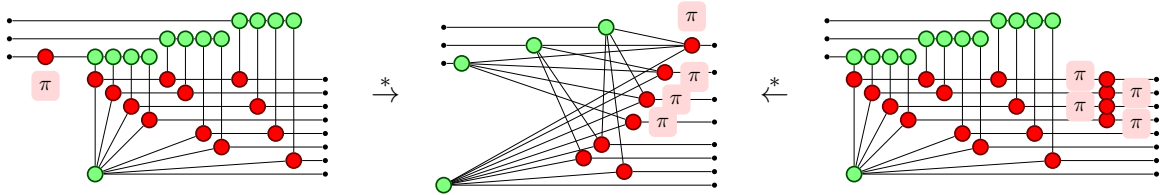
        \begin{minipage}{1.4\linewidth}
          \[
          \scalebox{0.8}{\begin{tikzpicture}[baseline={([yshift=-.5ex]current bounding box.center)}, x=0.3cm, y=0.3cm]
	\begin{pgfonlayer}{nodelayer}
		\node [style=wire] (0) at (6.5, -1.25) {};
		\node [style=Z] (1) at (2.75, 4.25) {};
		\node [style=wire] (2) at (6.5, -0.5) {};
		\node [style=X] (3) at (-5.25, 0.25) {};
		\node [style=X] (4) at (2.75, -0.5) {};
		\node [style=wire] (5) at (6.5, -2.75) {};
		\node [style=X] (6) at (-2.25, 1) {};
		\node [style=Z] (7) at (-2.25, 3.25) {};
		\node [style=X] (8) at (-3.25, -1.25) {};
		\node [style=wire] (9) at (-11, 3.25) {};
		\node [style=X] (10) at (-0.25, -2) {};
		\node [style=wire] (11) at (6.5, -4.25) {};
		\node [style=wire] (12) at (-11, 4.25) {};
		\node [style=Z] (13) at (-6.25, -4.25) {};
		\node [style=wire] (14) at (6.5, -3.5) {};
		\node [style=Z] (15) at (-3.25, 2.25) {};
		\node [style=Z] (16) at (4.75, 4.25) {};
		\node [style=X] (17) at (3.75, -2) {};
		\node [style=wire] (18) at (-11, 2.25) {};
		\node [style=Z] (19) at (-5.25, 2.25) {};
		\node [style=X] (20) at (-4.25, -0.5) {};
		\node [style=X] (21) at (-9, 2.25) {};
		\node [style=red angle below] (22) at (-9, 2.25) {$\pi$};
		\node [style=Z] (23) at (3.75, 4.25) {};
		\node [style=Z] (24) at (-6.25, 2.25) {};
		\node [style=X] (25) at (-1.25, 0.25) {};
		\node [style=wire] (26) at (6.5, -2) {};
		\node [style=Z] (27) at (0.75, 3.25) {};
		\node [style=X] (28) at (-6.25, 1) {};
		\node [style=X] (29) at (4.75, -3.5) {};
		\node [style=Z] (30) at (-4.25, 2.25) {};
		\node [style=Z] (31) at (-1.25, 3.25) {};
		\node [style=wire] (32) at (6.5, 0.25) {};
		\node [style=wire] (33) at (6.5, 1) {};
		\node [style=X] (34) at (1.75, 1) {};
		\node [style=Z] (35) at (1.75, 4.25) {};
		\node [style=Z] (36) at (-0.25, 3.25) {};
		\node [style=X] (37) at (0.75, -2.75) {};
	\end{pgfonlayer}
	\begin{pgfonlayer}{edgelayer}
		\draw (36) to (27);
		\draw (36) to (10);
		\draw (35) to (34);
		\draw (35) to (1);
		\draw (34) to (33);
		\draw (25) to (32);
		\draw (31) to (36);
		\draw (31) to (25);
		\draw (30) to (20);
		\draw (30) to (15);
		\draw (28) to (6);
		\draw (27) to (37);
		\draw (17) to (26);
		\draw (24) to (28);
		\draw (21) to (24);
		\draw (24) to (19);
		\draw (23) to (17);
		\draw (23) to (16);
		\draw (20) to (4);
		\draw (19) to (30);
		\draw (19) to (3);
		\draw (18) to (21);
		\draw (16) to (29);
		\draw (15) to (8);
		\draw (29) to (14);
		\draw (13) to (37);
		\draw (13) to (29);
		\draw (13) to (28);
		\draw (13) to (20);
		\draw (13) to (10);
		\draw (13) to (8);
		\draw (13) to (3);
		\draw (12) to (35);
		\draw (13) to (11);
		\draw (10) to (17);
		\draw (9) to (7);
		\draw (7) to (31);
		\draw (7) to (6);
		\draw (6) to (34);
		\draw (37) to (5);
		\draw (3) to (25);
		\draw (4) to (2);
		\draw (1) to (23);
		\draw (1) to (4);
		\draw (8) to (0);
	\end{pgfonlayer}
\end{tikzpicture}}
          \quad \stackrel{*}{\to} \quad
          \scalebox{0.8}{\begin{tikzpicture}[baseline={([yshift=-.5ex]current bounding box.center)}, x=0.3cm, y=0.3cm]
	\begin{pgfonlayer}{nodelayer}
		\node [style=wire] (0) at (10, -1.25) {};
		\node [style=Z] (1) at (4, 4) {};
		\node [style=wire] (2) at (10, 0) {};
		\node [style=wire] (3) at (10, -3.25) {};
		\node [style=X] (4) at (6.25, 0) {};
		\node [style=red angle] (5) at (6.25, 0) {$\pi$};
		\node [style=wire] (6) at (-5, 3) {};
		\node [style=X] (7) at (7.25, 1.5) {};
		\node [style=red angle] (8) at (7.25, 1.25) {$\pi$};
		\node [style=wire] (9) at (10, -4.75) {};
		\node [style=wire] (10) at (-5, 4) {};
		\node [style=Z] (11) at (-5, -4.75) {};
		\node [style=wire] (12) at (10, -4) {};
		\node [style=wire] (13) at (-5, 2) {};
		\node [style=Z] (14) at (-4, 2) {};
		\node [style=X] (15) at (8.75, 3) {};
		\node [style=red angle above] (16) at (8.75, 3) {$\pi$};
		\node [style=wire] (17) at (10, -2.5) {};
		\node [style=X] (18) at (5, -4) {};
		\node [style=Z] (19) at (0, 3) {};
		\node [style=wire] (20) at (10, 1.5) {};
		\node [style=wire] (21) at (10, 3) {};
		\node [style=X] (22) at (3.75, -2.5) {};
		\node [style=X] (23) at (2.75, -3.25) {};
		\node [style=X] (24) at (5.5, -1.25) {};
		\node [style=red angle] (25) at (5.5, -1.25) {$\pi$};
	\end{pgfonlayer}
	\begin{pgfonlayer}{edgelayer}
		\draw (21) to (15);
		\draw (20) to (7);
		\draw (19) to (23);
		\draw (19) to (22);
		\draw (19) to (15);
		\draw (19) to (7);
		\draw (17) to (22);
		\draw (14) to (24);
		\draw (14) to (15);
		\draw (14) to (7);
		\draw (14) to (4);
		\draw (14) to (13);
		\draw (18) to (12);
		\draw (11) to (24);
		\draw (11) to (23);
		\draw (11) to (22);
		\draw (11) to (18);
		\draw (11) to (15);
		\draw (11) to (7);
		\draw (11) to (4);
		\draw (1) to (10);
		\draw (11) to (9);
		\draw (19) to (6);
		\draw (23) to (3);
		\draw (2) to (4);
		\draw (1) to (22);
		\draw (1) to (18);
		\draw (1) to (15);
		\draw (1) to (4);
		\draw (0) to (24);
	\end{pgfonlayer}
\end{tikzpicture}}
          \quad \stackrel{*}{\leftarrow} \quad
          \scalebox{0.8}{\begin{tikzpicture}[baseline={([yshift=-.5ex]current bounding box.center)}, x=0.3cm, y=0.3cm]
	\begin{pgfonlayer}{nodelayer}
		\node [style=X] (0) at (7.25, -1.25) {};
		\node [style=red angle] (1) at (7.25, -1.45) {$\pi$};
		\node [style=X] (2) at (7.25, 0.25) {};
		\node [style=red angle] (3) at (7.25, 0.25) {$\pi$};
		\node [style=X] (4) at (7.25, 1) {};
		\node [style=red angle left] (5) at (7.25, 1.2) {$\pi$};
		\node [style=X] (6) at (7.25, -0.5) {};
		\node [style=red angle left] (7) at (7.25, -0.5) {$\pi$};
		\node [style=wire] (8) at (10, 1) {};
		\node [style=Z] (9) at (2.25, 4.25) {};
		\node [style=wire] (10) at (10, 0.25) {};
		\node [style=X] (11) at (-5.75, 0.25) {};
		\node [style=X] (12) at (2.25, -0.5) {};
		\node [style=wire] (13) at (10, -2.75) {};
		\node [style=X] (14) at (-2.75, 1) {};
		\node [style=Z] (15) at (-2.75, 3.25) {};
		\node [style=X] (16) at (-3.75, -1.25) {};
		\node [style=wire] (17) at (-8.25, 3.25) {};
		\node [style=X] (18) at (-0.75, -2) {};
		\node [style=wire] (19) at (10, -4.25) {};
		\node [style=wire] (20) at (-8.25, 4.25) {};
		\node [style=Z] (21) at (-6.75, -4.25) {};
		\node [style=wire] (22) at (10, -3.5) {};
		\node [style=Z] (23) at (-3.75, 2.25) {};
		\node [style=Z] (24) at (4.25, 4.25) {};
		\node [style=X] (25) at (3.25, -2) {};
		\node [style=wire] (26) at (-8.25, 2.25) {};
		\node [style=Z] (27) at (-5.75, 2.25) {};
		\node [style=X] (28) at (-4.75, -0.5) {};
		\node [style=Z] (29) at (3.25, 4.25) {};
		\node [style=Z] (30) at (-6.75, 2.25) {};
		\node [style=X] (31) at (-1.75, 0.25) {};
		\node [style=wire] (32) at (10, -2) {};
		\node [style=Z] (33) at (0.25, 3.25) {};
		\node [style=X] (34) at (-6.75, 1) {};
		\node [style=X] (35) at (4.25, -3.5) {};
		\node [style=Z] (36) at (-4.75, 2.25) {};
		\node [style=Z] (37) at (-1.75, 3.25) {};
		\node [style=wire] (38) at (10, -0.5) {};
		\node [style=wire] (39) at (10, -1.25) {};
		\node [style=X] (40) at (1.25, 1) {};
		\node [style=Z] (41) at (1.25, 4.25) {};
		\node [style=Z] (42) at (-0.75, 3.25) {};
		\node [style=X] (43) at (0.25, -2.75) {};
	\end{pgfonlayer}
	\begin{pgfonlayer}{edgelayer}
		\draw (42) to (33);
		\draw (42) to (18);
		\draw (41) to (40);
		\draw (41) to (9);
		\draw (40) to (4);
		\draw (0) to (39);
		\draw (6) to (38);
		\draw (37) to (42);
		\draw (37) to (31);
		\draw (36) to (28);
		\draw (36) to (23);
		\draw (34) to (14);
		\draw (33) to (43);
		\draw (25) to (32);
		\draw (31) to (2);
		\draw (30) to (34);
		\draw (30) to (27);
		\draw (29) to (25);
		\draw (29) to (24);
		\draw (28) to (12);
		\draw (27) to (36);
		\draw (27) to (11);
		\draw (26) to (30);
		\draw (24) to (35);
		\draw (23) to (16);
		\draw (35) to (22);
		\draw (21) to (43);
		\draw (21) to (35);
		\draw (21) to (34);
		\draw (21) to (28);
		\draw (21) to (18);
		\draw (21) to (16);
		\draw (21) to (11);
		\draw (20) to (41);
		\draw (21) to (19);
		\draw (18) to (25);
		\draw (17) to (15);
		\draw (16) to (0);
		\draw (15) to (37);
		\draw (15) to (14);
		\draw (14) to (40);
		\draw (43) to (13);
		\draw (12) to (6);
		\draw (11) to (31);
		\draw (2) to (10);
		\draw (9) to (29);
		\draw (9) to (12);
		\draw (4) to (8);
	\end{pgfonlayer}
\end{tikzpicture}}
          \]
        \end{minipage}}
    \end{center}
    The right-hand rewrite sequence is purely reduction, while the
    left-hand sequence makes use of one expansion step.  Both parts of
    the proof can be accomplished by the simproc
    \texttt{push\_pauli\_x}, shown in Figure \ref{fig:pushpaulix}.
    The proof has a total of 32 individual steps.  The other equations
    are proved in the same way, although the proofs of equations
    \eqref{eq:z1}, \eqref{eq:z2}, and \eqref{eq:z3} use a
    \texttt{push\_pauli\_z} simproc, which is the colour dual of
    \texttt{push\_pauli\_x}.
\end{proof}
\end{proposition}

\begin{corollary}\label{cor:codewordbasis}
  The computational basis states are prepared as shown in Table \ref{tab:basis}.
\end{corollary}

\subsection{The CNOT}
\label{sec:cnot}

Since the CNOT gate preserves the computational basis states, by
inspecting Table \ref{tab:basis} we can work out what the encoded CNOT
should be.  For example, let:
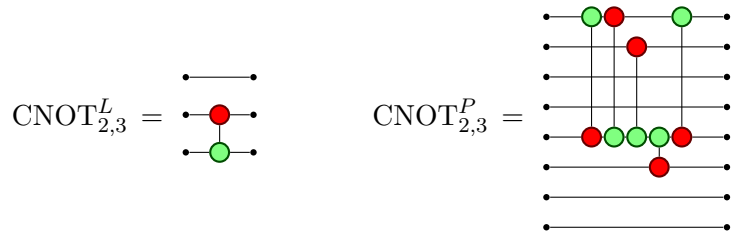
\[
\mathrm{CNOT}_{2,3}^L \;=\;\; \begin{tikzpicture}[baseline={([yshift=-.5ex]current bounding box.center)}, x=0.3cm, y=0.5cm]
	\begin{pgfonlayer}{nodelayer}
		\node [style=wire] (0) at (-1.5, 1) {};
		\node [style=wire] (1) at (-1.5, 0) {};
		\node [style=wire] (2) at (1.5, 1) {};
		\node [style=wire] (3) at (1.5, -1) {};
		\node [style=wire] (4) at (1.5, 0) {};
		\node [style=Z] (5) at (0, -1) {};
		\node [style=X] (6) at (0, 0) {};
		\node [style=wire] (7) at (-1.5, -1) {};
	\end{pgfonlayer}
	\begin{pgfonlayer}{edgelayer}
		\draw [style=simple] (5) to (6);
		\draw [style=simple] (6) to (4);
		\draw [style=simple] (1) to (6);
		\draw [style=simple] (0) to (2);
		\draw [style=simple] (7) to (5);
		\draw [style=simple] (5) to (3);
	\end{pgfonlayer}
\end{tikzpicture}
\qquad
\qquad
\mathrm{CNOT}_{2,3}^P \;=\;\; \begin{tikzpicture}[baseline={([yshift=-.5ex]current bounding box.center)}, x=0.3cm, y=0.4cm]
	\begin{pgfonlayer}{nodelayer}
		\node [style=wire] (b6) at (-8.0, -1.75) {};
		\node [style=wire] (b5) at (-8.0, -0.75) {};
		\node [style=wire] (b8) at (0.0, -1.75) {};
		\node [style=wire] (b15) at (0.0, 1.25) {};
		\node [style=X] (v4) at (-4.0, 3.25) {};
		\node [style=wire] (b13) at (0.0, -2.75) {};
		\node [style=wire] (b14) at (0.0, 0.25) {};
		\node [style=wire] (b1) at (-8.0, 3.25) {};
		\node [style=wire] (b10) at (0.0, 3.25) {};
		\node [style=Z] (v52) at (-6.0, 4.25) {};
		\node [style=wire] (b0) at (-8.0, 4.25) {};
		\node [style=wire] (b11) at (0.0, 4.25) {};
		\node [style=wire] (b9) at (0.0, -0.75) {};
		\node [style=Z] (v3) at (-3.0, 0.25) {};
		\node [style=wire] (b2) at (-8.0, 2.25) {};
		\node [style=Z] (v1) at (-5.0, 0.25) {};
		\node [style=X] (v0) at (-5.0, 4.25) {};
		\node [style=wire] (b7) at (-8.0, -2.75) {};
		\node [style=Z] (v7) at (-2.0, 4.25) {};
		\node [style=Z] (v2) at (-4.0, 0.25) {};
		\node [style=X] (v53) at (-6.0, 0.25) {};
		\node [style=wire] (b12) at (0.0, 2.25) {};
		\node [style=X] (v5) at (-3.0, -0.75) {};
		\node [style=wire] (b4) at (-8.0, 0.25) {};
		\node [style=wire] (b3) at (-8.0, 1.25) {};
		\node [style=X] (v6) at (-2.0, 0.25) {};
	\end{pgfonlayer}
	\begin{pgfonlayer}{edgelayer}
		\draw [style=simple] (v7) to (v6);
		\draw [style=simple] (b3) to (b15);
		\draw [style=simple] (b4) to (v53);
		\draw [style=simple] (v2) to (v3);
		\draw [style=simple] (v4) to (v2);
		\draw [style=simple] (b7) to (b13);
		\draw [style=simple] (v0) to (v7);
		\draw [style=simple] (v0) to (v1);
		\draw [style=simple] (v52) to (v0);
		\draw [style=simple] (v53) to (v1);
		\draw [style=simple] (v1) to (v2);
		\draw [style=simple] (b2) to (b12);
		\draw [style=simple] (v3) to (v6);
		\draw [style=simple] (v3) to (v5);
		\draw [style=simple] (v5) to (b9);
		\draw [style=simple] (v7) to (b11);
		\draw [style=simple] (b0) to (v52);
		\draw [style=simple] (v52) to (v53);
		\draw [style=simple] (v4) to (b10);
		\draw [style=simple] (b1) to (v4);
		\draw [style=simple] (v6) to (b14);
		\draw [style=simple] (b5) to (v5);
		\draw [style=simple] (b6) to (b8);
	\end{pgfonlayer}
\end{tikzpicture}

\]
\begin{proposition}\label{prop:cnot}
  With the above definitions, $\mathrm{CNOT}_{2,3}^L = \Enc^\dag \circ
  \mathrm{CNOT}_{2,3}^P \circ \Enc$.
  \begin{proof}
    We reduce as shown:
    
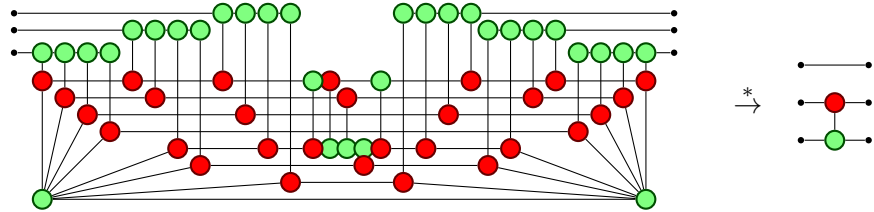
\[
    \begin{tikzpicture}[baseline={([yshift=-.5ex]current bounding box.center)}, x=0.3cm, y=0.3cm]
	\begin{pgfonlayer}{nodelayer}
		\node [style=X] (v28) at (8.25, 2.0) {};
		\node [style=Z] (v45) at (9.25, 3.25) {};
		\node [style=Z] (v9) at (-5.5, 5.0) {};
		\node [style=X] (v14) at (-13.5, 1.25) {};
		\node [style=Z] (v58) at (0.5, 2.0) {};
		\node [style=X] (v22) at (-5.5, 0.5) {};
		\node [style=Z] (v42) at (1.5, 5.0) {};
		\node [style=X] (v54) at (-1.75, 2.0) {};
		\node [style=Z] (v39) at (12.25, 3.25) {};
		\node [style=X] (v31) at (6.25, -1.0) {};
		\node [style=Z] (v41) at (7.25, 4.25) {};
		\node [style=Z] (v50) at (-1.75, -1.0) {};
		\node [style=X] (v17) at (-10.5, 2.0) {};
		\node [style=Z] (v4) at (-10.5, 4.25) {};
		\node [style=Z] (v38) at (4.5, 5.0) {};
		\node [style=wire] (b14) at (13.5, 4.25) {};
		\node [style=X] (v26) at (11.25, 1.25) {};
		\node [style=X] (v16) at (-11.5, -0.25) {};
		\node [style=wire] (b1) at (-15.75, 4.25) {};
		\node [style=X] (v19) at (-8.5, -1.0) {};
		\node [style=Z] (v33) at (11.25, 3.25) {};
		\node [style=X] (v40) at (7.25, 1.25) {};
		\node [style=Z] (v52) at (-2.5, 2.0) {};
		\node [style=wire] (b0) at (-15.75, 5.0) {};
		\node [style=Z] (v12) at (-14.5, -3.25) {};
		\node [style=X] (v30) at (9.25, -0.25) {};
		\node [style=X] (v27) at (3.5, 0.5) {};
		\node [style=X] (v43) at (1.5, -2.5) {};
		\node [style=Z] (v3) at (-11.5, 3.25) {};
		\node [style=Z] (v11) at (-3.5, 5.0) {};
		\node [style=wire] (b16) at (13.5, 5.0) {};
		\node [style=X] (v23) at (-4.5, -1.0) {};
		\node [style=wire] (b2) at (-15.75, 3.25) {};
		\node [style=X] (v35) at (2.5, -1.0) {};
		\node [style=wire] (b18) at (13.5, 3.25) {};
		\node [style=Z] (v32) at (2.5, 5.0) {};
		\node [style=Z] (v1) at (-13.5, 3.25) {};
		\node [style=X] (v15) at (-12.5, 0.5) {};
		\node [style=Z] (v56) at (-1.0, -1.0) {};
		\node [style=Z] (v25) at (5.25, 4.25) {};
		\node [style=Z] (v57) at (-0.25, -1.0) {};
		\node [style=Z] (v10) at (-4.5, 5.0) {};
		\node [style=Z] (v0) at (-14.5, 3.25) {};
		\node [style=X] (v18) at (-9.5, 1.25) {};
		\node [style=X] (v44) at (12.25, 2.0) {};
		\node [style=Z] (v47) at (6.25, 4.25) {};
		\node [style=X] (v51) at (-2.5, -1.0) {};
		\node [style=Z] (v7) at (-7.5, 4.25) {};
		\node [style=Z] (v34) at (3.5, 5.0) {};
		\node [style=X] (v13) at (-14.5, 2.0) {};
		\node [style=X] (v24) at (-3.5, -2.5) {};
		\node [style=X] (v46) at (4.5, 2.0) {};
		\node [style=Z] (v2) at (-12.5, 3.25) {};
		\node [style=X] (v37) at (10.25, 0.5) {};
		\node [style=X] (v53) at (-0.25, -1.75) {};
		\node [style=X] (v49) at (5.25, -1.75) {};
		\node [style=Z] (v36) at (12.25, -3.25) {};
		\node [style=Z] (v5) at (-9.5, 4.25) {};
		\node [style=X] (v55) at (-1.0, 1.25) {};
		\node [style=X] (v21) at (-6.5, 2.0) {};
		\node [style=Z] (v8) at (-6.5, 5.0) {};
		\node [style=Z] (v6) at (-8.5, 4.25) {};
		\node [style=X] (v20) at (-7.5, -1.75) {};
		\node [style=Z] (v48) at (8.25, 4.25) {};
		\node [style=Z] (v29) at (10.25, 3.25) {};
		\node [style=X] (v75) at (0.5, -1.0) {};
		\node [style=none] (padl) at (-16.75, -4.25) {};
		\node [style=none] (padr) at (14.5, 6.0) {};
		\node [style=none] (padu) at (-16.75, 6.0) {};
		\node [style=none] (padd) at (14.5, -4.25) {};
	\end{pgfonlayer}
	\begin{pgfonlayer}{edgelayer}
		\draw [style=simple] (v29) to (v37);
		\draw [style=simple] (v29) to (v33);
		\draw [style=simple] (v45) to (v29);
		\draw [style=simple] (v20) to (v53);
		\draw [style=simple] (v6) to (v7);
		\draw [style=simple] (v6) to (v19);
		\draw [style=simple] (v8) to (v21);
		\draw [style=simple] (v8) to (v9);
		\draw [style=simple] (v21) to (v52);
		\draw [style=simple] (v55) to (v56);
		\draw [style=simple] (v5) to (v6);
		\draw [style=simple] (v5) to (v18);
		\draw [style=simple] (v49) to (v36);
		\draw [style=simple] (v37) to (v36);
		\draw [style=simple] (v44) to (v36);
		\draw [style=simple] (v43) to (v36);
		\draw [style=simple] (v53) to (v49);
		\draw [style=simple] (v57) to (v53);
		\draw [style=simple] (v2) to (v15);
		\draw [style=simple] (v2) to (v3);
		\draw [style=simple] (v58) to (v46);
		\draw [style=simple] (v24) to (v43);
		\draw [style=simple] (v13) to (v17);
		\draw [style=simple] (v34) to (v38);
		\draw [style=simple] (v7) to (v20);
		\draw [style=simple] (v52) to (v51);
		\draw [style=simple] (v18) to (v55);
		\draw [style=simple] (v0) to (v13);
		\draw [style=simple] (v0) to (v1);
		\draw [style=simple] (v10) to (v23);
		\draw [style=simple] (v10) to (v11);
		\draw [style=simple] (v57) to (v75);
		\draw [style=simple] (v25) to (v49);
		\draw [style=simple] (v25) to (v47);
		\draw [style=simple] (v56) to (v57);
		\draw [style=simple] (v15) to (v22);
		\draw [style=simple] (v1) to (v2);
		\draw [style=simple] (v1) to (v14);
		\draw [style=simple] (v32) to (v34);
		\draw [style=simple] (v32) to (v35);
		\draw [style=simple] (v42) to (v32);
		\draw [style=simple] (v39) to (b18);
		\draw [style=simple] (v75) to (v35);
		\draw [style=simple] (b2) to (v0);
		\draw [style=simple] (v23) to (v51);
		\draw [style=simple] (v38) to (b16);
		\draw [style=simple] (v11) to (v24);
		\draw [style=simple] (v3) to (v16);
		\draw [style=simple] (v37) to (v27);
		\draw [style=simple] (v34) to (v27);
		\draw [style=simple] (v30) to (v36);
		\draw [style=simple] (v30) to (v45);
		\draw [style=simple] (v12) to (v20);
		\draw [style=simple] (v12) to (v36);
		\draw [style=simple] (v12) to (v24);
		\draw [style=simple] (v12) to (v13);
		\draw [style=simple] (v12) to (v15);
		\draw [style=simple] (v12) to (v19);
		\draw [style=simple] (v12) to (v16);
		\draw [style=simple] (v12) to (v14);
		\draw [style=simple] (b0) to (v8);
		\draw [style=simple] (v52) to (v54);
		\draw [style=simple] (v55) to (v40);
		\draw [style=simple] (v41) to (v40);
		\draw [style=simple] (v33) to (v39);
		\draw [style=simple] (v19) to (v23);
		\draw [style=simple] (b1) to (v4);
		\draw [style=simple] (v16) to (v30);
		\draw [style=simple] (v26) to (v36);
		\draw [style=simple] (v40) to (v26);
		\draw [style=simple] (v33) to (v26);
		\draw [style=simple] (v48) to (b14);
		\draw [style=simple] (v38) to (v46);
		\draw [style=simple] (v4) to (v5);
		\draw [style=simple] (v4) to (v17);
		\draw [style=simple] (v17) to (v21);
		\draw [style=simple] (v51) to (v50);
		\draw [style=simple] (v50) to (v56);
		\draw [style=simple] (v54) to (v50);
		\draw [style=simple] (v41) to (v48);
		\draw [style=simple] (v47) to (v41);
		\draw [style=simple] (v31) to (v36);
		\draw [style=simple] (v47) to (v31);
		\draw [style=simple] (v35) to (v31);
		\draw [style=simple] (v39) to (v44);
		\draw [style=simple] (v54) to (v58);
		\draw [style=simple] (v42) to (v43);
		\draw [style=simple] (v22) to (v27);
		\draw [style=simple] (v58) to (v75);
		\draw [style=simple] (v14) to (v18);
		\draw [style=simple] (v9) to (v10);
		\draw [style=simple] (v9) to (v22);
		\draw [style=simple] (v48) to (v28);
		\draw [style=simple] (v46) to (v28);
		\draw [style=simple] (v28) to (v44);
	\end{pgfonlayer}
\end{tikzpicture}

    \quad
    \stackrel{*}{\to}
    \quad
    \begin{tikzpicture}[baseline={([yshift=-.5ex]current bounding box.center)}, x=0.3cm, y=0.5cm]
	\begin{pgfonlayer}{nodelayer}
		\node [style=wire] (0) at (-1.5, 1) {};
		\node [style=wire] (1) at (-1.5, 0) {};
		\node [style=wire] (2) at (1.5, 1) {};
		\node [style=wire] (3) at (1.5, -1) {};
		\node [style=wire] (4) at (1.5, 0) {};
		\node [style=Z] (5) at (0, -1) {};
		\node [style=X] (6) at (0, 0) {};
		\node [style=wire] (7) at (-1.5, -1) {};
	\end{pgfonlayer}
	\begin{pgfonlayer}{edgelayer}
		\draw [style=simple] (5) to (6);
		\draw [style=simple] (6) to (4);
		\draw [style=simple] (1) to (6);
		\draw [style=simple] (0) to (2);
		\draw [style=simple] (7) to (5);
		\draw [style=simple] (5) to (3);
	\end{pgfonlayer}
\end{tikzpicture}
    \]
    Noticing that neither diagram has any angles, we can again employ
    \texttt{rotate\_simp}; this does most of the work, but terminates
    in a different quasi-normal form to what we want.  However, it
    requires only one expansion step to get over this hurdle, and
    simple reductions can automatically finish the proof.  The
    complete proof has 75 steps, and can be found in the Supplementary Material
    \ref{SUPP-sec:cnot}.
  \end{proof}
\end{proposition}

The other possible CNOTs can be discovered by applying permutations to
the above, and the same proof technique will establish their
correctness. 

\subsection{The CCZ}
\label{sec:ccz}

As noted in Section \ref{sec:code}, the main point of interest of this
code is the fact that the CCZ gate can be implemented fault tolerantly
using only $T$ and $T^\dag$ operations on the physical qubits.
Precisely:
\[
\mathrm{CCZ}_{1,2,3}^P = 
T \otimes T^{\dagger} \otimes T^{\dagger} \otimes T \otimes 
T^{\dagger} \otimes T \otimes T \otimes T^{\dagger}\,.
\]
To establish the correctness of this claim, ideally we would prove:

\begin{equation}
  \mathrm{CCZ}_{1,2,3}^L = \Enc^\dag \circ \mathrm{CCZ}_{1,2,3}^P\circ
  \Enc\;,\label{eq:ccz}
\end{equation}
or, in pictures:

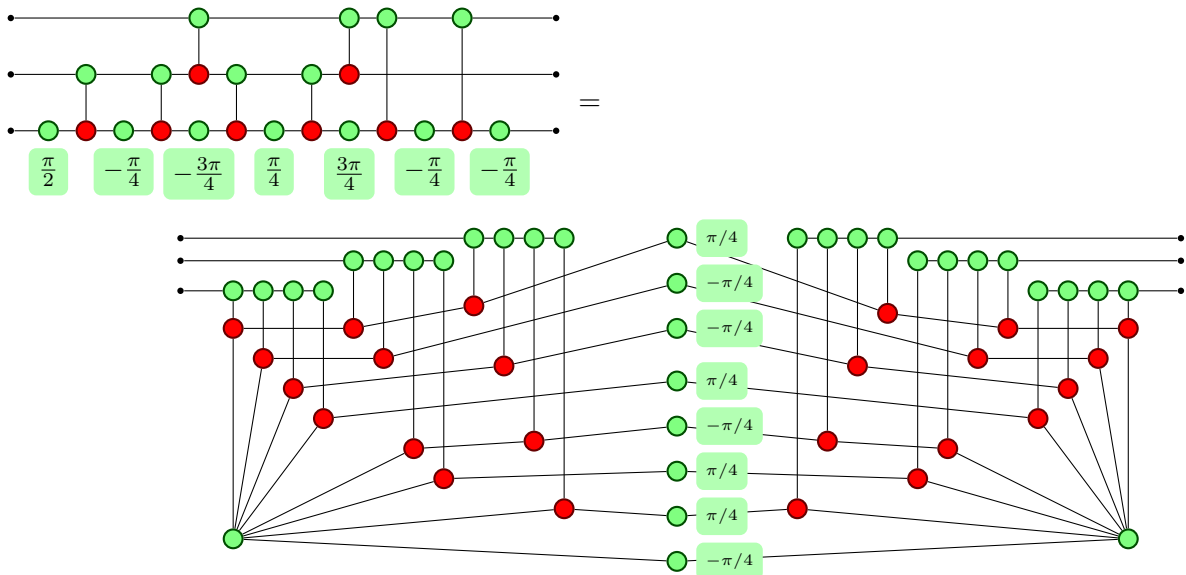
\begin{center}
  \makebox[\textwidth][c]{
    \begin{minipage}[c]{1.5\textwidth}
\[
\beginpgfgraphicnamed{ccz-circuit}
\InputIfFileExists{ccz-circuit.tikz}{}{\input{./figures/ccz-circuit.tikz}}
\endpgfgraphicnamed \;\; =
        \qquad\qquad\qquad\qquad\qquad\qquad
\qquad\qquad\qquad\qquad\qquad
\]
\[
\qquad\qquad        
\beginpgfgraphicnamed{fault-toler-ops/CCZ/CCZ}
\begin{tikzpicture}[baseline={([yshift=-.5ex]current bounding box.center)}, x=0.4cm, y=0.4cm]
	\begin{pgfonlayer}{nodelayer}
		\node [style=Z] (0) at (-0.5, 1) {};
		\node [style=green angle] (1) at (-0.5, 1) {$\scriptscriptstyle-\pi/4$};
		\node [style=Z] (2) at (8.5, 3.25) {};
		\node [style=Z] (3) at (-6.25, 4) {};
		\node [style=wire] (4) at (16.25, 2.25) {};
		\node [style=X] (5) at (-14.25, 0) {};
		\node [style=X] (6) at (13.5, 0) {};
		\node [style=X] (7) at (-6.25, -0.25) {};
		\node [style=Z] (8) at (5.5, 4) {};
		\node [style=X] (9) at (9.5, 0) {};
		\node [style=Z] (10) at (14.5, -6) {};
		\node [style=X] (11) at (6.5, 1.5) {};
		\node [style=Z] (12) at (-0.5, -3.75) {};
		\node [style=green angle] (13) at (-0.5, -3.75) {$\scriptscriptstyle\pi/4$};
		\node [style=X] (14) at (14.5, 1) {};
		\node [style=Z] (15) at (4.5, 4) {};
		\node [style=Z] (16) at (13.5, 2.25) {};
		\node [style=X] (17) at (-11.25, 1) {};
		\node [style=Z] (18) at (-11.25, 3.25) {};
		\node [style=X] (19) at (5.5, -0.25) {};
		\node [style=Z] (20) at (-0.5, 4) {};
		\node [style=green angle] (21) at (-0.5, 4) {$\scriptscriptstyle\pi/4$};
		\node [style=X] (22) at (-12.25, -2) {};
		\node [style=wire] (23) at (-17, 3.25) {};
		\node [style=X] (24) at (-9.25, -3) {};
		\node [style=Z] (25) at (-0.5, -6.75) {};
		\node [style=green angle] (26) at (-0.5, -6.75) {$\scriptscriptstyle-\pi/4$};
		\node [style=Z] (27) at (3.5, 4) {};
		\node [style=X] (28) at (7.5, -4) {};
		\node [style=wire] (29) at (-17, 4) {};
		\node [style=Z] (30) at (-15.25, -6) {};
		\node [style=Z] (31) at (-0.5, -2.25) {};
		\node [style=green angle] (32) at (-0.5, -2.25) {$\scriptscriptstyle-\pi/4$};
		\node [style=Z] (33) at (-0.5, 2.5) {};
		\node [style=green angle] (34) at (-0.5, 2.5) {$\scriptscriptstyle-\pi/4$};
		\node [style=Z] (35) at (6.5, 4) {};
		\node [style=Z] (36) at (-12.25, 2.25) {};
		\node [style=Z] (37) at (-4.25, 4) {};
		\node [style=X] (38) at (-5.25, -2.75) {};
		\node [style=wire] (39) at (-17, 2.25) {};
		\node [style=Z] (40) at (-0.5, -5.25) {};
		\node [style=green angle] (41) at (-0.5, -5.25) {$\scriptscriptstyle\pi/4$};
		\node [style=Z] (42) at (-14.25, 2.25) {};
		\node [style=X] (43) at (-13.25, -1) {};
		\node [style=X] (44) at (11.5, -2) {};
		\node [style=X] (45) at (12.5, -1) {};
		\node [style=Z] (46) at (-5.25, 4) {};
		\node [style=Z] (47) at (-15.25, 2.25) {};
		\node [style=X] (48) at (-10.25, 0) {};
		\node [style=Z] (49) at (7.5, 3.25) {};
		\node [style=Z] (50) at (10.5, 3.25) {};
		\node [style=Z] (51) at (14.5, 2.25) {};
		\node [style=Z] (52) at (-8.25, 3.25) {};
		\node [style=X] (53) at (-15.25, 1) {};
		\node [style=X] (54) at (-4.25, -5) {};
		\node [style=Z] (55) at (9.5, 3.25) {};
		\node [style=Z] (56) at (-13.25, 2.25) {};
		\node [style=X] (57) at (4.5, -2.75) {};
		\node [style=X] (58) at (8.5, -3) {};
		\node [style=Z] (59) at (12.5, 2.25) {};
		\node [style=X] (60) at (3.5, -5) {};
		\node [style=Z] (61) at (-10.25, 3.25) {};
		\node [style=wire] (62) at (16.25, 3.25) {};
		\node [style=wire] (63) at (16.25, 4) {};
		\node [style=X] (64) at (10.5, 1) {};
		\node [style=X] (65) at (-7.25, 1.75) {};
		\node [style=Z] (66) at (-7.25, 4) {};
		\node [style=Z] (67) at (-9.25, 3.25) {};
		\node [style=X] (68) at (-8.25, -4) {};
		\node [style=Z] (69) at (11.5, 2.25) {};
		\node [style=Z] (70) at (-0.5, -0.75) {};
		\node [style=green angle] (71) at (-0.5, -0.75) {$\scriptscriptstyle\pi/4$};
	\end{pgfonlayer}
	\begin{pgfonlayer}{edgelayer}
		\draw (70) to (44);
		\draw (69) to (59);
		\draw (44) to (69);
		\draw (68) to (12);
		\draw (67) to (52);
		\draw (67) to (24);
		\draw (66) to (65);
		\draw (66) to (3);
		\draw (65) to (20);
		\draw (14) to (64);
		\draw (35) to (63);
		\draw (50) to (62);
		\draw (61) to (67);
		\draw (61) to (48);
		\draw (27) to (60);
		\draw (10) to (60);
		\draw (59) to (45);
		\draw (59) to (16);
		\draw (58) to (10);
		\draw (57) to (58);
		\draw (15) to (57);
		\draw (56) to (43);
		\draw (56) to (36);
		\draw (55) to (50);
		\draw (55) to (9);
		\draw (54) to (40);
		\draw (53) to (17);
		\draw (52) to (68);
		\draw (51) to (14);
		\draw (50) to (64);
		\draw (49) to (28);
		\draw (49) to (2);
		\draw (48) to (33);
		\draw (47) to (53);
		\draw (47) to (42);
		\draw (46) to (38);
		\draw (46) to (37);
		\draw (45) to (10);
		\draw (44) to (10);
		\draw (43) to (7);
		\draw (42) to (56);
		\draw (42) to (5);
		\draw (40) to (60);
		\draw (39) to (47);
		\draw (38) to (31);
		\draw (37) to (54);
		\draw (36) to (22);
		\draw (33) to (9);
		\draw (31) to (57);
		\draw (30) to (68);
		\draw (30) to (54);
		\draw (30) to (53);
		\draw (30) to (43);
		\draw (30) to (25);
		\draw (30) to (24);
		\draw (30) to (22);
		\draw (30) to (5);
		\draw (29) to (66);
		\draw (28) to (10);
		\draw (27) to (15);
		\draw (25) to (10);
		\draw (24) to (38);
		\draw (23) to (18);
		\draw (22) to (70);
		\draw (20) to (11);
		\draw (45) to (19);
		\draw (8) to (19);
		\draw (18) to (61);
		\draw (18) to (17);
		\draw (17) to (65);
		\draw (16) to (51);
		\draw (6) to (16);
		\draw (15) to (8);
		\draw (14) to (10);
		\draw (12) to (28);
		\draw (64) to (11);
		\draw (35) to (11);
		\draw (9) to (6);
		\draw (8) to (35);
		\draw (7) to (0);
		\draw (6) to (10);
		\draw (5) to (48);
		\draw (51) to (4);
		\draw (3) to (46);
		\draw (3) to (7);
		\draw (2) to (58);
		\draw (2) to (55);
		\draw (0) to (19);
	\end{pgfonlayer}
\end{tikzpicture}}
\endpgfgraphicnamed
\]
    \end{minipage}}
\end{center}

Unlike our previous results, the non-Clifford $T$ gate plays a crucial
role here, and the \zxcalculus is known to be incomplete for the
Clifford+T fragment of quantum mechanics, even after the addition of
equations which we do not consider here \cite{Perdrix:2015aa}.  It
remains possible that \eqref{eq:ccz} may be established from the usual
\zxcalculus axioms, however this proved beyond our reach.  We resort
to brute force:

\begin{proposition}\label{prop:ccz-correctness}
  For $x,y \in \{0,1\}$ such that $xy = 0$ we have:
  \begin{gather*}
    (\Enc^\dag \circ \mathrm{CCZ}_{1,2,3}^P\circ\Enc) \ket{11+} = 
    \ket{11-}\\
    (\Enc^\dag \circ \mathrm{CCZ}_{1,2,3}^P\circ\Enc) \ket{11-} =
    \ket{11+}\\    
    (\Enc^\dag \circ \mathrm{CCZ}_{1,2,3}^P\circ\Enc) \ket{xy\pm} = \ket{xy\pm}
  \end{gather*}
  \begin{proof}
    We treat each of the 8 cases separately by performing, for
    example, the reduction:
    \[
    \scalebox{0.9}
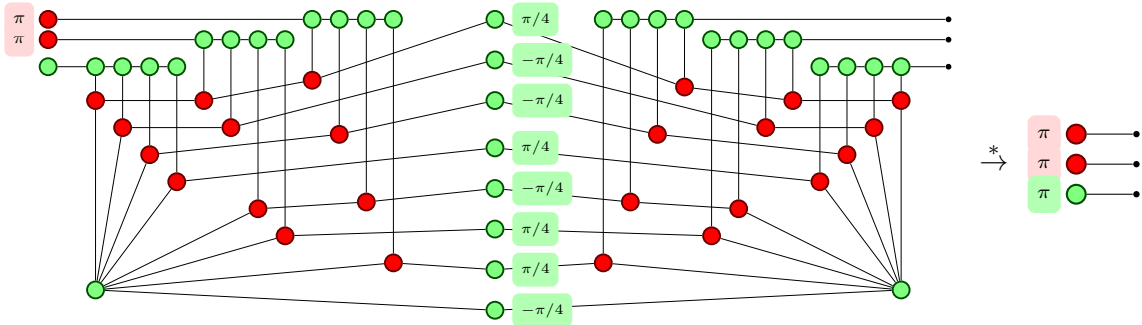
{\begin{tikzpicture}[baseline={([yshift=-.5ex]current bounding box.center)}, x=0.4cm, y=0.4cm]
	\begin{pgfonlayer}{nodelayer}
		\node [style=X] (23) at (-17, 3.25) {};
    \node [style=red angle left] (l23) at (-17, 3.25) {$\scriptstyle\pi$};
		\node [style=X] (29) at (-17, 4) {};
    \node [style=red angle left] (l29) at (-17,4) {$\scriptstyle\pi$};
		\node [style=Z] (39) at (-17, 2.25) {};

		\node [style=Z] (0) at (-0.5, 1) {};
		\node [style=green angle] (1) at (-0.5, 1) {$\scriptscriptstyle-\pi/4$};
		\node [style=Z] (2) at (8.5, 3.25) {};
		\node [style=Z] (3) at (-6.25, 4) {};
		\node [style=wire] (4) at (16.25, 2.25) {};
		\node [style=X] (5) at (-14.25, 0) {};
		\node [style=X] (6) at (13.5, 0) {};
		\node [style=X] (7) at (-6.25, -0.25) {};
		\node [style=Z] (8) at (5.5, 4) {};
		\node [style=X] (9) at (9.5, 0) {};
		\node [style=Z] (10) at (14.5, -6) {};
		\node [style=X] (11) at (6.5, 1.5) {};
		\node [style=Z] (12) at (-0.5, -3.75) {};
		\node [style=green angle] (13) at (-0.5, -3.75) {$\scriptscriptstyle\pi/4$};
		\node [style=X] (14) at (14.5, 1) {};
		\node [style=Z] (15) at (4.5, 4) {};
		\node [style=Z] (16) at (13.5, 2.25) {};
		\node [style=X] (17) at (-11.25, 1) {};
		\node [style=Z] (18) at (-11.25, 3.25) {};
		\node [style=X] (19) at (5.5, -0.25) {};
		\node [style=Z] (20) at (-0.5, 4) {};
		\node [style=green angle] (21) at (-0.5, 4) {$\scriptscriptstyle\pi/4$};
		\node [style=X] (22) at (-12.25, -2) {};

		\node [style=X] (24) at (-9.25, -3) {};
		\node [style=Z] (25) at (-0.5, -6.75) {};
		\node [style=green angle] (26) at (-0.5, -6.75) {$\scriptscriptstyle-\pi/4$};
		\node [style=Z] (27) at (3.5, 4) {};
		\node [style=X] (28) at (7.5, -4) {};

		\node [style=Z] (30) at (-15.25, -6) {};
		\node [style=Z] (31) at (-0.5, -2.25) {};
		\node [style=green angle] (32) at (-0.5, -2.25) {$\scriptscriptstyle-\pi/4$};
		\node [style=Z] (33) at (-0.5, 2.5) {};
		\node [style=green angle] (34) at (-0.5, 2.5) {$\scriptscriptstyle-\pi/4$};
		\node [style=Z] (35) at (6.5, 4) {};
		\node [style=Z] (36) at (-12.25, 2.25) {};
		\node [style=Z] (37) at (-4.25, 4) {};
		\node [style=X] (38) at (-5.25, -2.75) {};

		\node [style=Z] (40) at (-0.5, -5.25) {};
		\node [style=green angle] (41) at (-0.5, -5.25) {$\scriptscriptstyle\pi/4$};
		\node [style=Z] (42) at (-14.25, 2.25) {};
		\node [style=X] (43) at (-13.25, -1) {};
		\node [style=X] (44) at (11.5, -2) {};
		\node [style=X] (45) at (12.5, -1) {};
		\node [style=Z] (46) at (-5.25, 4) {};
		\node [style=Z] (47) at (-15.25, 2.25) {};
		\node [style=X] (48) at (-10.25, 0) {};
		\node [style=Z] (49) at (7.5, 3.25) {};
		\node [style=Z] (50) at (10.5, 3.25) {};
		\node [style=Z] (51) at (14.5, 2.25) {};
		\node [style=Z] (52) at (-8.25, 3.25) {};
		\node [style=X] (53) at (-15.25, 1) {};
		\node [style=X] (54) at (-4.25, -5) {};
		\node [style=Z] (55) at (9.5, 3.25) {};
		\node [style=Z] (56) at (-13.25, 2.25) {};
		\node [style=X] (57) at (4.5, -2.75) {};
		\node [style=X] (58) at (8.5, -3) {};
		\node [style=Z] (59) at (12.5, 2.25) {};
		\node [style=X] (60) at (3.5, -5) {};
		\node [style=Z] (61) at (-10.25, 3.25) {};
		\node [style=wire] (62) at (16.25, 3.25) {};
		\node [style=wire] (63) at (16.25, 4) {};
		\node [style=X] (64) at (10.5, 1) {};
		\node [style=X] (65) at (-7.25, 1.75) {};
		\node [style=Z] (66) at (-7.25, 4) {};
		\node [style=Z] (67) at (-9.25, 3.25) {};
		\node [style=X] (68) at (-8.25, -4) {};
		\node [style=Z] (69) at (11.5, 2.25) {};
		\node [style=Z] (70) at (-0.5, -0.75) {};
		\node [style=green angle] (71) at (-0.5, -0.75) {$\scriptscriptstyle\pi/4$};
	\end{pgfonlayer}
	\begin{pgfonlayer}{edgelayer}
		\draw (70) to (44);
		\draw (69) to (59);
		\draw (44) to (69);
		\draw (68) to (12);
		\draw (67) to (52);
		\draw (67) to (24);
		\draw (66) to (65);
		\draw (66) to (3);
		\draw (65) to (20);
		\draw (14) to (64);
		\draw (35) to (63);
		\draw (50) to (62);
		\draw (61) to (67);
		\draw (61) to (48);
		\draw (27) to (60);
		\draw (10) to (60);
		\draw (59) to (45);
		\draw (59) to (16);
		\draw (58) to (10);
		\draw (57) to (58);
		\draw (15) to (57);
		\draw (56) to (43);
		\draw (56) to (36);
		\draw (55) to (50);
		\draw (55) to (9);
		\draw (54) to (40);
		\draw (53) to (17);
		\draw (52) to (68);
		\draw (51) to (14);
		\draw (50) to (64);
		\draw (49) to (28);
		\draw (49) to (2);
		\draw (48) to (33);
		\draw (47) to (53);
		\draw (47) to (42);
		\draw (46) to (38);
		\draw (46) to (37);
		\draw (45) to (10);
		\draw (44) to (10);
		\draw (43) to (7);
		\draw (42) to (56);
		\draw (42) to (5);
		\draw (40) to (60);
		\draw (39) to (47);
		\draw (38) to (31);
		\draw (37) to (54);
		\draw (36) to (22);
		\draw (33) to (9);
		\draw (31) to (57);
		\draw (30) to (68);
		\draw (30) to (54);
		\draw (30) to (53);
		\draw (30) to (43);
		\draw (30) to (25);
		\draw (30) to (24);
		\draw (30) to (22);
		\draw (30) to (5);
		\draw (29) to (66);
		\draw (28) to (10);
		\draw (27) to (15);
		\draw (25) to (10);
		\draw (24) to (38);
		\draw (23) to (18);
		\draw (22) to (70);
		\draw (20) to (11);
		\draw (45) to (19);
		\draw (8) to (19);
		\draw (18) to (61);
		\draw (18) to (17);
		\draw (17) to (65);
		\draw (16) to (51);
		\draw (6) to (16);
		\draw (15) to (8);
		\draw (14) to (10);
		\draw (12) to (28);
		\draw (64) to (11);
		\draw (35) to (11);
		\draw (9) to (6);
		\draw (8) to (35);
		\draw (7) to (0);
		\draw (6) to (10);
		\draw (5) to (48);
		\draw (51) to (4);
		\draw (3) to (46);
		\draw (3) to (7);
		\draw (2) to (58);
		\draw (2) to (55);
		\draw (0) to (19);
	\end{pgfonlayer}
\end{tikzpicture}}
    \;\;\stackrel{*}{\to}\;\;
    \begin{tikzpicture}[baseline={([yshift=-.5ex]current bounding box.center)}, x=0.4cm, y=0.4cm]
	\begin{pgfonlayer}{nodelayer}
		\node [style=X] (23) at (0, 3) {};
    \node [style=red angle left] (l23) at (0, 3) {$\scriptstyle\pi$};
		\node [style=X] (29) at (0, 4) {};
    \node [style=red angle left] (l29) at (0,4) {$\scriptstyle\pi$};
		\node [style=Z] (39) at (0, 2) {};
    \node [style=green angle left] (l29) at (0,2) {$\scriptstyle\pi$};
		\node [style=wire] (1) at (2, 3) {};
		\node [style=wire] (2) at (2, 4) {};
		\node [style=wire] (3) at (2, 2) {};
	\end{pgfonlayer}
	\begin{pgfonlayer}{edgelayer}
  \draw (23) to (1);
  \draw (29) to (2);
  \draw (39) to (3);
	\end{pgfonlayer}
\end{tikzpicture}
    \]
    A typical such proof has 80--90 steps; the majority can be
    performed by either the basic simplifier or \texttt{rotate\_simp},
    however at several points specially prepared lemmas are required
    to simplify specific graphs that occurred during the proof.  The
    proof strategy did not appear amenable to automation.  The
    complete proof is in Supplementary Material \ref{SUPP-sec:ccz}.
  \end{proof}
\end{proposition}

\begin{corollary} \label{cor:ccz-works}
  $\mathrm{CCZ}_{1,2,3}^P$ acts as a logical CCZ on code
  words\footnote{Evaluating the map on a basis is not sufficient to
    establish the equality unless global phase is taken into account.
    However evaluating the map on the equal sum of the basis elements will
    suffice to complete the proof, even without phases.  This is done in
    the electronic material.
}.
\end{corollary}



\section{Conclusions and Future Work}
\label{sec:future-work}

In this paper we have formalised the Smallest Interesting Colour Code
in the \zxcalculus, and provided formal proofs certifying
some of its basic properties.  Along the way we discovered a novel
quantum circuit for encoding three input qubits into the 8-qubit
codeword space.  We did not consider a deterministic decoder, nor the
error detecting circuit for this code.

The size of the graphs involved in these terms, and the number of
steps involved in the proofs, meant that Quantomatic was indispensible
to this work.  The power of the standard simplifiers
(\texttt{basic\_simp} and especially \texttt{rotate\_simp}) is
striking; given such a hammer, everything looks like a nail.
Therefore, while lemmas could in principle be used to shorten and add
conceptual structure to proofs\footnote{Indeed, while typesetting the
  enormous proof of Proposition \ref{prop:ccz-correctness} in 
  Supplementary Material \ref{sec:ccz} we discovered a much shorter and clearer
  proof; sadly too late to be included in this version of the
  paper.}, 
in this work they have mostly been
used to climb out of a position where the simplifiers got stuck.  From
this point of view there are two main uses of lemmas: (1) to chain
together some rewrites in an order the simplifier would not choose;
and (2) to embed a rewrite in a larger graph and thus restrict where
the matcher may apply it.  For example we count
many variations on the theme of ``unspidering'' among the frequently
employed lemmas.  Lacking any other mechanism to control where
rewrites will be applied, lemmas of the second type are essential to
writing useful simprocs.  Care must also be taken with the formulation
of the axioms to control matching.  For example,
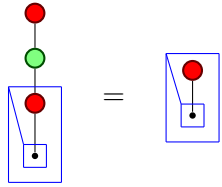
\texttt{gen\_bialg.qrule} (See Appendix \ref{sec:ruleset}) is
formulated too generally to be useful.  While it does encapsulate the
general form of the bialgbera equation, its left-hand-side also
matches on the empty graph and every edge of a 2-coloured graph.  The
resulting deluge of trivial matches is more hindrance than help.

During our attack on Proposition
\ref{prop:ccz-correctness} we codified Selinger and Bian's
generators and relations for 2-qubit Clifford+T circuits
\cite{Selinger2016Relations-for-2} hoping the additional equations
would help. However, their novel
relations involve large circuits, which did not occur in the
graphs we were working with, and so could not
be used.  A useful rule must have a left-hand-side which captures some
non-trivial feature of a term already in reduced form (\ie
two-colured and simple).

The most glaring absence from this work is the Clifford group.  No
treatment was attempted, due to limited available time, but is the
most obvious next step.  Beyond that, this code is an interesting test
case for further study of the Clifford+T fragment in the \zxcalculus.
Recall from Section \ref{sec:ccz} that the encoded CCZ is very simple
in comparison to the usual 3-qubit circuit.  It is also rather simpler
than the encoded CNOT circuit we presented in Section \ref{sec:cnot},
which, without performing any optimisation, contains five physical
CNOT operations.  Since CCZ is Clifford equivalent to Toffoli, the
CNOT can be generated by CCZs and Cliffords.  This implies there
exists a family of equations between CNOT circuits and Clifford+T
circuits which we conjecture are not provable in the \zxcalculus.
This offers a new approach to completing the calculus, while helping
with applications along the way.  More generally, applying automated
reasoning and the \zxcalculus to verify practical gate synthesis
protocols such as
\cite{Campbell2016Unifying-gate-s,PhysRevA.95.022316} is an important
area for further investigation.

\small
\bibliography{all}

\newpage
\normalsize



\appendix

\section{Electronic Resources}
\label{sec:electronic-resources}

All the work described in this paper is available as a downloadable
quantomatic project from the following URL.
\begin{center}
  \url{https://gitlab.cis.strath.ac.uk/kwb13215/Colour-Code-QPL}
\end{center}
Please download it and play around!

\section{The Ruleset}
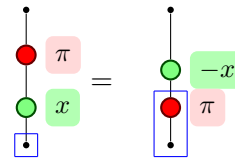
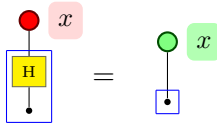
\label{sec:ruleset}

\newcommand{\showrule}[1]{
  \begin{minipage}[t]{0.3\linewidth}
    \texttt{\detokenize{#1}.qrule} \\
    \ctikzfig{figures/ruleset/#1}
  \end{minipage}}

\tikzset{every picture/.style={x=0.4cm,y=0.4cm}}
\tikzset{font={\footnotesize}}
\tikzstyle{bbox}=[] 

\showrule{green_sp}
\showrule{green_id}
\showrule{green_loop}

\noindent
\showrule{green_scalar}
\showrule{green_pi}
\showrule{green_copy}

\noindent
\showrule{green_elim}
\showrule{red_sp}
\showrule{red_id}

\noindent
\showrule{red_loop}
\showrule{red_scalar}
\showrule{red_pi}

\noindent
\showrule{red_copy}
\showrule{red_elim}
\showrule{gen_bialg}

\noindent
\showrule{gen_bialg_simp}
\showrule{hopf}
\showrule{euler}

\noindent
\showrule{euler2}
\showrule{h-cancel}
\showrule{green_to_red}

\noindent
\showrule{red_to_green}


\section{Proofs omitted from main article}
\label{sec:proofs-omitted-from}

For those readers who prefer to see the proofs written down rather
than as a Quantomatic proof development, they are available in
supplementary document.  This available in two places:
\begin{itemize}
\item The second author's website, at the time of writing:\\
  \url{http://personal.strath.ac.uk/ross.duncan/}
\item The \emph{first} version of this paper stored on the arXiv
  (arxiv:1706.02717v1): \\
  \url{https://arxiv.org/abs/1706.02717v1}
\end{itemize}

\end{document}